\documentclass[twocolumn]{jpsj2} 
\def\gsim{\mathop {\vtop {\ialign {##\crcr 
$\hfil \displaystyle {>}\hfil $\crcr \noalign {\kern1pt \nointerlineskip } 
$\,\sim$ \crcr \noalign {\kern1pt}}}}\limits}
\def\lsim{\mathop {\vtop {\ialign {##\crcr 
$\hfil \displaystyle {<}\hfil $\crcr \noalign {\kern1pt \nointerlineskip } 
$\,\,\sim$ \crcr \noalign {\kern1pt}}}}\limits}

\title{
Valence Fluctuations Revealed by Magnetic Field and Pressure Scans: 
Comparison with Experiments in 
YbXCu$_4$ (X=In, Ag, Cd) and CeYIn$_5$ (Y=Ir, Rh) 
}

\author{Shinji \textsc{Watanabe}, 
Atsushi \textsc{Tsuruta}$^1$, 
Kazumasa \textsc{Miyake}$^1$,
and 
Jacques \textsc{Flouquet}$^2$
}

\inst{Department of Applied Physics, University of Tokyo, Hongo 7-3-1, 
Bunkyo-ku, Tokyo, 113-8656, Japan
\\
Division of Materials Physics, Department of Materials Engineering Science, Graduate School of
Engineering Science, Osaka University, Toyonaka, Osaka 560-8531, Japan$^1$
\\
D{\'e}partement de la Recherche Fondamentale sur la Mati{\`e}re Condense{\'e}, SPSMS, CEA Grenoble,
17 rue des Martyrs, 38054 Grenoble Cedex 9, France$^2$
}

\abst{
The mechanism of how critical end points of the first-order valence transition (FOVT) are controlled 
by a magnetic field is discussed. 
We demonstrate that 
critical temperature is suppressed to be a quantum critical point (QCP) by a magnetic field. 
This results explain 
the field dependence of the isostructural FOVT 
observed in Ce metal and $\rm YbInCu_4$. 
Magnetic field scan can make the system reenter in a critical valence fluctuation region. 
Even in intermediate-valence materials, the QCP 
is induced by applying a magnetic field, at which magnetic susceptibility also diverges. 
The driving force of the field-induced QCP is shown to be a cooperative phenomenon of 
the Zeeman effect and the Kondo effect, which creates a distinct energy scale 
from the Kondo temperature. 
The key concept is that the closeness to the QCP of the FOVT is 
vital in understanding Ce- and Yb-based heavy-fermions. 
This explains the peculiar magnetic and transport responses in $\rm CeYIn_5$ (Y=Ir, Rh) 
and metamagnetic transition in $\rm YbXCu_4$ for X=In 
as well as the sharp contrast between X=Ag and Cd. 
}

\kword{quantum critical point, first-order valence transition, valence fluctuations, CeIrIn$_5$, 
CeRhIn$_5$, YbInCu$_4$, YbAgCu$_4$, YbCdCu$_4$}

\begin{document}
\maketitle

\section{Introduction}

Quantum critical phenomena in itinerant fermion systems with strong correlations 
have attracted much attention in condensed matter physics. 
When the continuous transition temperature of the magnetically ordered phase is 
suppressed by controlling material parameters and reaches absolute zero, 
the quantum critical point (QCP) emerges. 
In the paramagnetic metal phase near the QCP, enhanced spin fluctuations cause 
non-Fermi liquid behaviors in physical quantities exhibiting quantum criticalities~\cite{Moriya,Hertz,Millis}, 
and even trigger other instabilities such as unconventional superconductivity. 
So far, the magnetic QCP and the role of spin fluctuations have been extensively discussed 
from both theoretical~\cite{Moriya,Hertz,Millis} and experimental sides~\cite{Stewart}. 

\begin{figure}[h]
\includegraphics[width=75mm]{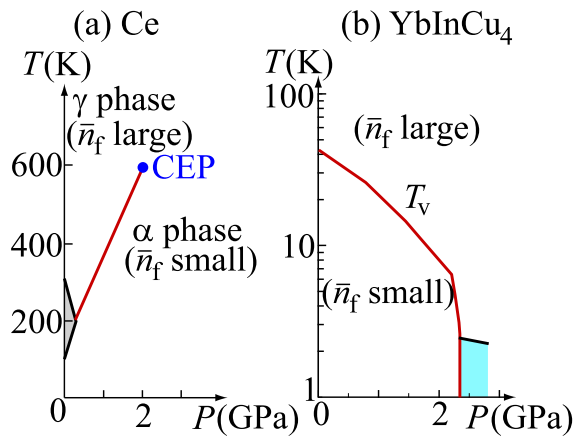}  
\caption{\label{fig:Ce}(color online) 
Temperature-pressure phase diagram of (a) Ce metal~\cite{Cevalence} and (b) YbInCu$_4$~\cite{Mito2003,Park2006PRL}. 
(a) The first-order valence transition between the $\gamma$-phase and the $\alpha$-phase (solid line) 
terminates at the critical end point (CEP) (solid circle) in the fcc lattice. 
The shaded area around $P=0$ represents the $\beta$-phase. 
(b) The first-order valence transition (solid line) is suppressed under pressure 
in the cubic AuBe$_5$ (C15b type) lattice. 
Note that the $T$ axis is shown on the logarithmic scale.
The shaded area represents the magnetically ordered phase~\cite{Mito2003}. 
$\bar{n}_{\rm f}$ denotes the number of electrons per Ce in (a) and 
the number of holes per Yb in (b) (see text). 
}
\end{figure}

Recently, critical phenomena associated with 
charge degrees of freedom have attracted attention. 
In particular, 
valence instability and its critical fluctuations have attracted much attention 
as a possible origin of anomalies in Ce- and Yb-based heavy-fermion systems~\cite{MNO,M07}.
Valence transition phenomena were closely studied four decades ago under the label of 
intermediate valence. 
Evidence of its occurrence comes from the $\gamma$-$\alpha$ transition 
of Ce metal~\cite{Bridgeman} six decades ago (see Fig.~\ref{fig:Ce}(a)) with 
first-order valence transition (FOVT), which in the (temperature $T$, pressure $P$) plane starts at 
$T_{\rm v}\sim 120$~K 
and terminates at the critical end point (CEP) ($T_{\rm CEP}\approx 600$~K, 
$P_{\rm CEP}\approx 2$~GPa)~\cite{Cevalence}. 
As the intercept of the $T_{\rm v}(P)$ line of the FOVT is rather high at $P=0$, 
no quantum criticality was discovered~\cite{Flouquet2006}.
For many anomalous Ce compounds, no FOVT has been detected 
despite the fact that their valence deviates from three where the occupation number $(\bar{n}_{\rm f})$ 
of the 4f shell is unity; the conditions are such that the system is always 
in a valence crossover regime, i.e., 
the system escapes from the CEP but, as will be stressed later, 
can feel its proximity.

An excellent example of FOVT for Yb systems was reported for YbInCu$_4$ 
(see Fig.~\ref{fig:Ce}(b))~\cite{Felner,Kojima1989,Sarrao1996PRB,Mito2003,Park2006PRL}, 
and the Yb case can be regarded 
as the hole analog of Ce; $\bar{n}_{\rm f}$ being the hole occupation number of the 4f shell 
(14 for Yb$^{2+}$ and 13 for Yb$^{3+}$ where $\bar{n}_{\rm f}=1$). 
Although YbInCu$_4$ as well as Ce metal is a prototypical example that shows the FOVT, 
most Ce- and Yb-based heavy-fermions seem to be in a valence-crossover regime. 
When the CEP is suppressed by tuning material parameters 
and enters the Fermi-degeneracy regime,  
diverging valence fluctuations are considered to be coupled with Fermi-surface instability. 
This multiple instability seems to give the key mechanism 
that dominates the low-temperature properties of materials 
including valence-fluctuating ions such as Ce and Yb. 
It may play a dominant role in heavy-fermion quantum instabilities. 

In Fig.~\ref{fig:Ce}(b), 
with increasing $P$, $T_{\rm v}(P)$ decreases and even becomes sufficiently low 
such that FOVT reaches 
the magnetic boundary~\cite{Park2006PRL,Mito2003} clearly in a narrow $P$ window. 
The interplay between valence transition and magnetic transition can be strong. 
Looking carefully into the disappearance under pressure of long range magnetism in 
Ce- and Yb-based heavy-fermion systems, this interplay is often strong~\cite{Flouquet2006}.
There are only a few material series like the $\rm CeCu_2Si_2$~\cite{jaccard,holms,fujiwara,yuan,yuan2} series 
and $\rm CeRu_2Si_2$~\cite{Flouquet2006} series
where the magnetic QCP at $P_{\rm c}$ is not coupled with valence fluctuation. 
Thus, discriminating between valence and spin quantum criticality is often difficult. 

The proof of valence fluctuations in the quantum degeneracy regime 
seems to be supported by evidence of the two-superconductivity mechanism 
in the $\rm CeCu_2Si_2$-$\rm CeCu_2Ge_2$ series 
where $P_{\rm v}-P_{\rm c}\sim 4$~GPa~\cite{jaccard,holms,fujiwara,yuan,yuan2}. 
A marked increase in the superconducting transition temperature 
$T_{\rm SC}$ is observed at a pressure where 
the valence of Ce changes sharply in $\rm CeCu_2Ge_2$~\cite{jaccard}, 
$\rm CeCu_2Si_2$~\cite{holms,fujiwara}, and $\rm CeCu_2(Ge_xSi_{1-x})_2$~\cite{yuan,yuan2}. 
The importance of quantum criticality is shown above $T_{\rm SC}$ by the observation of 
non-Fermi liquid $T$-linear resistivity in wide temperature region. 
The $T$-linear resistivity has been observed in a variety of Ce- and Yb-based 
heavy-fermion systems~\cite{M07}. 

Theoretically, the possibility of the valence-fluctuation-mediated superconductivity 
in the $P$-$T$ phase diagram of CeCu$_2$Ge$_2$ 
was pointed out in ref.~\citen{MNO}. 
It was shown that the $T$-linear resistivity emerges in a wide temperature range near the QCP of the valence transition~\cite{holms}. 
Residual resistivity was also shown to be markedly enhanced when the system is tuned to 
approach the QCP by controlling $P$ and/or the concentration of the chemical doping~\cite{MM}.
Near the QCP, the superconducting transition temperature was shown to be enhanced 
by valence fluctuations on the basis of 
 the slave-boson mean-field theory taking account of its Gaussian 
fluctuations~\cite{OM}. 
The stability and lattice-structure dependence of 
density-fluctuation-mediated superconductivity were 
argued phenomenologically~\cite{ML}. 
%
%

Recent numerical studies have revealed the significance of 
valence fluctuations near the QCP of 
the FOVT~\cite{WIM}  
and clarified its new aspects: 
The emergence of unconventional superconductivity 
due to an anomalous increase in the coherence of quasiparticles near the QCP, 
and the absence of phase separation as well as non diverging total charge compressibility 
even at the QCP at least in electronic origin 
due to the non conserving order parameter of the 
valence transition~\cite{WIM}.

In $(T, P)$ phase diagrams of heavy-fermion systems, magnetic, valence, and superconductivity 
boundaries can seriously cause interference. 
This suggests the idea that this interplay also occurs in 
the ($T$, magnetic field $h$) plane 
for $P$ close to $P_{\rm v}$ near the QEP. 
Valence fluctuations are essentially relative charge fluctuations 
between f and conduction electrons. 
Hence, 
it is highly nontrivial how magnetic field affects the valence QCP as well as QEP. 
To resolve these fundamental issues, we have theoretically studied 
the magnetic field dependence of 
the critical points of the FOVT~\cite{WTMF}.

In this paper, we report the mechanism of how the QCP as well as the CEP of the FOVT 
is controlled by a magnetic field in great detail. 
We discuss how this newly clarified mechanism gives an explanation 
of unresolved observations in Ce- and Yb-based systems. 
First, 
we show how critical end temperature is suppressed to absolute zero 
by applying a magnetic field, which explains the field dependence of 
the isostructural FOVT temperature 
observed in Ce metal and YbInCu$_4$. 
Our results also explain the peculiar magnetic response in CeIrIn$_5$, 
where 
the first-order transition line emerges in the temperature-magnetic field phase 
diagram, giving rise to the increase in residual resistivity 
as well as the appearance of the $T$-linear resistivity. 
The differences in the location of the material with respect to CEP 
explains the sharp contrast between YbAgCu$_4$ and YbCdCu$_4$ 
in their magnetization curves 
in spite of the fact that both have nearly the same Kondo temperatures. 
Our results indicate that the QCP as well as the FOVT 
is induced even in moderately intermediate valence materials by applying a magnetic field, 
which causes various anomalies such as non-Fermi liquid behavior in the resistivity, 
the increase in the residual resisitivity, 
and diverging magnetic susceptibility. 
We discuss the significance of the proximity to the critical points 
of the FOVT 
to understand unresolved phenomena in Ce- and Yb-based heavy-fermions. 
The key concept is the closeness to the QCP of the FOVT.

\section{First-Order Valence Transition 
under Magnetic Field at Finite Temperature}

\begin{figure}[h]
\includegraphics[width=75mm]{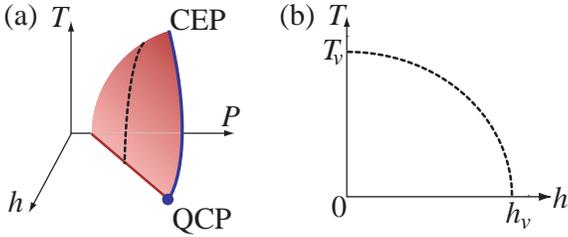}
\caption{\label{fig:TPH}(color online) (a) Schematic phase diagram, 
showing the FOVT surface in the $T$-$P$-$h$ space, where $P$ represents 
a control parameter (e.g., pressure and chemical concentration). 
The critical end points (CEPs) form a continuous transition line 
that reaches $T=0$ as the QCP. 
(b) FOVT line $(h/h_{\rm v})^2+(T/T_{\rm v})^2=1$~\cite{Dzero2,magCe,Immer,Basu} 
in the $T$-$h$ plane for a fixed $P$, corresponding to the dashed line 
in (a). 
}
\end{figure}

To give a quantitative outlook of the field dependence of the valence transition, 
let us consider the Claudius-Clapeyron 
relation for the FOVT temperature $T_{\rm v}$: 
\begin{eqnarray}
\frac{d T_{\rm v}}{d h}=-\frac{m_{\rm K}-m_{\rm MV}}{S_{\rm K}-S_{\rm MV}}, 
\label{eq:CCR}
\end{eqnarray}
where $m$ and $S$ denote the magnetization and entropy, respectively, 
and $h$ denotes the magnetic field. 
Here, K indicates the Kondo regime where the f-electron (hole) density 
per site $\bar{n}_{\rm f}$ 
is close to 1 in the Ce (Yb)-based system, i.e., $\rm Ce^{3+} (4f^1)$ 
and $\rm Yb^{3+} (4f^{13})$, 
and MV indicates the ``mixed-valence" regime with $\bar{n}_{\rm f} <1$~\cite{defMV}. 
Since the magnetization as well as the entropy in the Kondo regime 
is larger than those in the MV regime, 
as observed in the specific heat and the uniform susceptibility, 
it turns out that 
$T_{\rm v}$ is suppressed by applying $h$ (see Fig.~\ref{fig:TPH}(a)). 
Then, the critical end point is eventually suppressed to $T=0$~K by $h$. 

Furthermore, the field dependence of $T_{\rm v}$ in the zero-temperature limit 
is also derived using the above relation: 
For $T\to 0$, the entropy shows the $T$-linear behavior 
in both the Kondo and MV regimes 
so that 
$S_{\rm K}-S_{\rm MV}$ is approximated to be proportional to $T_{\rm v}$ 
in the case where $T_{\rm v}$ is smaller than the characteristic energy scales 
in the Kondo and MV regimes. 
Noting that $m_{\rm K}-m_{\rm MV}$ is temperature-independent for $T\to 0$, we have 
$\delta T_{\rm v}/\delta h=-C_1/T_{\rm v}$, 
leading to $T_{\rm v}=\sqrt{2C_1}\sqrt{h_{\rm v}-h}$ with constants $C_1$, 
which explains well the observed behavior in the Ce metal~\cite{magCe} 
and $\rm YbInCu_4$~\cite{Immer} (see Fig.~\ref{fig:TPH}(b)). 
We stress here that our analysis not only provides a firm ground for small-$T_{\rm v}$ behavior
by considering the coherence of electrons essential for low temperature, 
but also interpolates the high $T_{\rm v}$ 
satisfying the relation $(h/h_{\rm v})^2+(T/T_{\rm v})^2=1$~\cite{Dzero2,magCe,Immer,Basu} to zero temperature, 
since this relation was derived by assuming an isolated atomic entropy~\cite{magCe},
which is justified only in the high-temperature regime. 
Although the above discussion is about the FOVT $T_{\rm v}$ temperature, 
it turns out that the critical end temperature $T_{\rm CEP}$ is also suppressed, 
as shown in Figs.~\ref{fig:TPH}(a) and \ref{fig:TPH}(b).

According to the Maxwell relation, 
the volume variation $V_{\rm o}(h)$ with $h$ is related to the pressure derivative 
of the magnetization $m$ $(\partial V_{\rm o}/\partial h=-\partial m/\partial P)$ 
proportional to the Pauli susceptibility in this paramagnetic state, and 
thus inversely proportional to its Kondo temperature. 
For Ce, $\partial m/\partial P$ decreases under $P$ as $\bar{n}_{\rm f}$; 
thus, $\partial V_{\rm o}/\partial h$ is positive 
and the system enters in the trivalent state upon increasing $h$. 
The same occurs for Yb 
because, now, $\partial m/\partial P$ increases 
with $P$; thus, $n_{\rm f}$: $\partial V_{\rm o}/\partial h$ is negative. 
The phenomenological dependence of $T_{\rm v}(h)$ as reported in Fig.~\ref{fig:TPH}(a) 
was derived in agreement with this picture~\cite{Dzero2}.


\section{Extended Periodic Anderson Model}

Although we have shown that the $h$ dependence of the critical temperature $T_{\rm CEP}$ 
can be understood from the viewpoint of the free-energy gain induced by the larger entropy 
in the Kondo regime, 
it is highly nontrivial how the QCP of the FOVT 
is controlled by $h$ at $T=0$. 
To proceed with our analysis, 
we introduce a minimal model that describes the essential part 
of Ce- and Yb-based systems~\cite{Falikov,Varma,FK,Hewson,HF1,Zlatic,Golstev}: 
\begin{eqnarray}
H&=& \sum_{{\bf k}\sigma}\varepsilon_{\bf k}
c_{{\bf k}\sigma}^{\dagger}c_{{\bf k}\sigma}
+\varepsilon_{\rm f}\sum_{i\sigma}n^{\rm f}_{i\sigma}
\nonumber \\
&+&V\sum_{i\sigma}\left(
f_{i\sigma}^{\dagger}c_{i\sigma}+c_{i\sigma}^{\dagger}f_{i\sigma}
\right)
+U\sum_{i=1}^{N}n_{i\uparrow}^{\rm f}n_{i\downarrow}^{\rm f}
\nonumber
\\
&+&U_{\rm fc}\sum_{i=1}^{N}n_{i}^{\rm f}n_{i}^{\rm c}
-h\sum_{i}(S_i^{{\rm f}z}+S_i^{{\rm c}z}), 
\label{eq:PAM}
\end{eqnarray}
where 
$c_{i\sigma}$ $(c^{\dagger}_{i\sigma})$ is the annihilation (creation) operator 
of the conduction electron at the $i$-th site with a spin $\sigma$, 
and $f_{i\sigma}$ $(f^{\dagger}_{i\sigma})$ is that of the f electron. 
The number operator is defined by 
$n^{\rm a}_{i\sigma}=a_{i\sigma}^{\dagger}a_{i\sigma}$ and 
$n^{\rm a}_{i}=n^{\rm a}_{i\uparrow}+n^{\rm a}_{i\downarrow}$ for ${\rm a}={\rm f}$ and c.  
Here, $\varepsilon_{\bf k}$ denotes the energy dispersion for conduction electrons. 
$\varepsilon_{\rm f}$ is the f level and $V$ is the hybridization between the f and conduction 
electrons. 
The effect of applying pressure is to increase both the hybridization $V$ and 
the f-level $\varepsilon_{\rm f}$ relatively to the Fermi level, the latter of which 
plays a more crucial role in the valence transition than the former 
in Ce and Yb compounds. 
In other words, the increase in pressure is parameterized essentially by 
that in $\varepsilon_{\rm f}$. 

The on-site Coulomb repulsion for f electrons is given by the $U$ term. 
The $U_{\rm fc}$ term is the Coulomb repulsion between the f and conduction electrons, 
which is considered to play an important role in the valence transition~\cite{Falikov,
Varma,FK,Hewson,HF1,Zlatic,Golstev,OM,WIM}. 
Namely, the periodic Anderson model without $U_{\rm fc}$ cannot explain a sharp or 
discontinuous valence transition as discussed in refs.~\citen{OM} and \citen{WIM}. 
For example, in the case of Ce metal that exhibits the $\gamma$-$\alpha$ transition, 
the 4f- and 5d-electron bands are located at the Fermi level~\cite{Ceband}.   
Since both 4f and 5d orbitals are located on the same Ce site, this $U_{\rm fc}$ term 
cannot be neglected. 
In Yb systems, $f_{i\sigma}$ $(f^{\dagger}_{i\sigma})$ 
is regarded as the annihilation (creation) operator of the f hole and hence 
$\varepsilon_{\rm f}$ denotes the f-hole level. 
For $\rm YbInCu_4$, a considerable magnitude of the In 5p and Yb 4f hybridization 
was pointed out by the band-structure calculation~\cite{Takegahara} 
and recent high-resolution photoemission spectra have detected a remarkable 
increase in the magnitude of the p-f hybridization at the FOVT~\cite{Yoshikawa}.
These results suggest the importance of both $V_{\bf k}$ and $U_{\rm fc}$. 

The reason why the critical-end temperature $T_{\rm CEP}$ is so high, 
that is as much as 600~K, 
in Ce metal in contrast to that in $\rm YbInCu_4$ can be understood in terms 
of the difference in the magnitude of $U_{\rm fc}$. 
In $\rm YbInCu_4$, $U_{\rm fc}$ originates from 
the intersite interaction, which should be 
smaller than that of Ce metal. 
This view also gives the reason why most Ce and Yb compounds 
only show valence crossover, but not FOVT. 
Namely, most of the compounds seem to have a moderate $U_{\rm fc}$ 
owing to its intersite origin, which is smaller than the critical value 
for causing a discontinuous jump of the valence. 
However, even in the valence-crossover regime, 
underlying effect of valence instability 
causes intriguing phenomena, as shown below. 
It should be noted that the importance of the $U_{\rm fc}$ term in playing a crucial role 
in the isostructural FOVT in $\rm YbInCu_4$
in the hole picture of eq.~(\ref{eq:PAM}) 
was pointed out in refs.~\citen{Zlatic} and~\citen{Golstev}. 
The last term in eq.~(\ref{eq:PAM}) is the Zeeman term with $h$ being 
the magnetic field including the g factors. 

\subsection{Slave-boson mean field theory}

We apply the slave-boson-mean-field theory~\cite{slbMF,OM} 
to the Hamiltonian eq.~(\ref{eq:PAM}) at $T=0$ where the 
slave-boson-mean-field theory is a reasonable approximation. 
To describe the state for $U=\infty$, 
we consider 
$Vf_{i\sigma}^{\dagger}b_{i}c_{i\sigma}$ 
instead of 
$V f_{i\sigma}^{\dagger}c_{i\sigma}$ in eq.~(\ref{eq:PAM}) 
by introducing the slave-boson operator $b_{i}$ at the $i$-th site 
to describe the ${\rm f}^{0}$ state 
and require the constraint 
$
\sum_{\sigma}n_{i\sigma}^{ \rm f}+b_{i}^{\dagger}b_{i}=1
$
by the method of the Lagrange multiplier 
$
\sum_{i}\lambda_{i}\left(
\sum_{\sigma}n_{i\sigma}^{ \rm f}+b_{i}^{\dagger}b_{i}-1
\right). 
$
For $H_{U_{\rm fc}}$ in eq.~(\ref{eq:PAM}), 
we employ mean-field decoupling as 
%
$
n_{i}^{ \rm f}n_{i}^{ c}\simeq 
n_{ \rm f}n_{i}^{ c}+n_{ c}n_{i}^{ \rm f}-\frac{1}{2}n_{ \rm f}n_{ c}. 
$
%
By approximating mean fields as uniform ones, i.e.,  
$b=\langle b_{i} \rangle$ and $\bar{\lambda}=\lambda_i$, 
the set of mean-field equations 
is obtained by  
${\partial\langle H \rangle}/{\partial \lambda}=0$ and 
${\partial \langle H \rangle}/{\partial b}=0$ 
as follows: 
\begin{eqnarray}
\bar{\lambda}
&=&
\frac{V^{2}}{N}
\sum_{{\bf k}\sigma}
\frac{f(E_{{\bf k}\sigma}^{-})-f(E_{{\bf k}\sigma}^{+})}
{\sqrt{(\bar{\varepsilon}_{{ \rm f}\sigma}
-\bar{\varepsilon}_{{\bf k}\sigma})^{2}+4{\bar{V}}^{2}}}, 
\\
1-
|\bar{b}|^{2}
&=&
\frac{1}{2N}
\sum_{{\bf k}\sigma,\pm}
\left[
1\pm\frac{\bar{\varepsilon}_{{ \rm f}\sigma}
-\bar{\varepsilon}_{{\bf k}\sigma}}
{\sqrt{(\bar{\varepsilon}_{{ \rm f}\sigma}
-\bar{\varepsilon}_{{\bf k}\sigma})^{2}+4\bar{V}^{2}}}
\right]
\nonumber
\\
& & \times f(E^{\pm}_{{\bf k}\sigma}),
\end{eqnarray}
and the following equation holds for the total electron number: 
\begin{eqnarray}
\bar{n}_{ \rm f}+\bar{n}_{ c}
=
\frac{1}{N}
\sum_{{\bf k}\sigma}
\left[
f(E^{-}_{{\bf k}\sigma})+f(E^{+}_{{\bf k}\sigma})
\right]. 
\end{eqnarray}
Here, $f(E)$ is the Fermi distribution function and 
$E^{\pm}_{{\bf k}\sigma}$ are the lower $(-)$ and upper $(+)$ 
hybridized bands for a quasiparticle with spin $\sigma$, respectively: 
\begin{eqnarray}
E^{\pm}_{{\bf k}\sigma}=
\frac{1}{2}
\left[
\bar{\varepsilon}_{{ \rm f}\sigma}
+\bar{\varepsilon}_{{\bf k}\sigma}
\pm
\sqrt{(
\bar{\varepsilon}_{{ \rm f}\sigma}
-\bar{\varepsilon}_{{\bf k}\sigma})^{2}+4\bar{V}^{2}}
\right], 
\end{eqnarray}
where 
$\bar{\varepsilon}_{{\bf k}\sigma}$, $\bar{\varepsilon}_{{ \rm f}\sigma}$, and $\bar{V}$ 
are defined by 
%
$
\bar{\varepsilon}_{{\bf k}\sigma}\equiv
\varepsilon_{\bf k}
+U_{ \rm fc}\bar{n}_{ \rm f}
-\frac{h\sigma}{2}
$, 
$
\bar{\varepsilon}_{{ \rm f}\sigma}\equiv
\varepsilon_{ \rm f}
+\bar{\lambda}
+U_{ \rm fc}\bar{n}_{ c}
-\frac{h\sigma}{2}
$ 
and 
$
\bar{V}\equiv
V|\bar{b}|
$. 
%
The dispersion of the conduction electrons is taken as 
$\varepsilon_{\bf k}={\bf k}^2/(2m)-D$ with $-D$ being set as 
the bottom of the conduction band 
and the density of states $N_{0}(\varepsilon)$ is set to satisfy the normalization 
condition, $\int_{-D}^{D}d\varepsilon N_{0}(\varepsilon)=1$ per spin 
in three dimension (see inset of Fig.~\ref{fig:h0PD}).  
We take $D$ as the energy unit and show the results for $V=0.5$ and the total filling 
$n=(\bar{n}_{\rm f}+\bar{n}_{{\rm c}})/2=7/8$ below. 

\subsection{Properties at zero magnetic field}

\begin{figure}
\includegraphics[width=85mm]{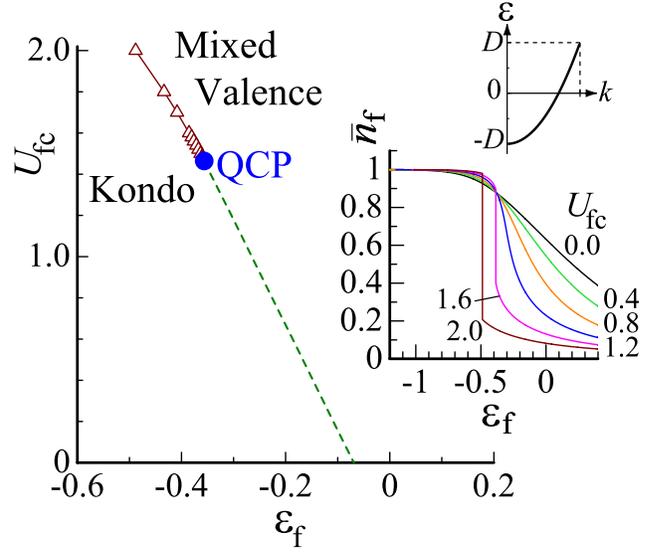}
\caption{\label{fig:h0PD}(color online) Ground-state phase diagram in the plane of $U_{\rm fc}$ and 
$\varepsilon_{\rm f}$ for $D=1$, $V=0.5$ at $n=7/8$. 
The FOVT line (solid line with open triangles) terminates at a QCP (a solid circle) for $h=0.00$. 
The dashed line represents the valence-crossover points at which $\chi_{\rm v}$ has a maximum 
as a function of $\varepsilon_{\rm f}$ for each $U_{\rm fc}$. 
The inset shows the energy band of conduction electrons, 
and $\bar{n}_{\rm f}$ vs $\varepsilon_{\rm f}$ for $U_{\rm fc}=0.0$, 0.4, 0.8, 1.2, 1.6, and 2.0 
under $h=0$. 
} 
\end{figure}

When $\varepsilon_{\rm f}$ is deep, the Kondo state with $\bar{n}_{\rm f}=1$ is realized. 
As $\varepsilon_{\rm f}$ increases, electrons move from the f level into the conduction band 
via hybridization, giving rise to the MV state. 
Hence, $\bar{n}_{\rm f}$ decreases gradually for $U_{\rm fc}=0$ 
as shown in the inset of Fig.~\ref{fig:h0PD}, as calculated using the slave-boson mean-field theory 
in model~(\ref{eq:PAM}). 
As $U_{\rm fc}$ increases, $\bar{n}_{\rm f}$ decreases sharply as a function of $\varepsilon_{\rm f}$. 
For large $U_{\rm fc}$, 
$\bar{n}_{\rm f}=\partial \langle H \rangle/\partial \varepsilon_{\rm f}$ 
shows a discontinuous jump, 
which indicates the level crossing of the ground states 
between the Kondo state and the MV state~\cite{OM,WIM,WTMF}. 
(see $U_{\rm fc}=1.6$ and 2.0 in the inset of Fig.~\ref{fig:h0PD}). 
The first-order transition is caused by $U_{\rm fc}$, 
since a large $U_{\rm fc}$ forces the electrons to pour into either the f level or 
the conduction band~\cite{WIM,WTMF}. 

\begin{figure}
\includegraphics[width=80mm]{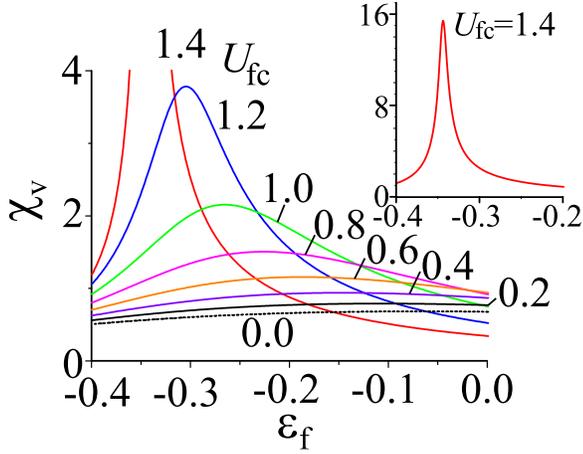}
\caption{\label{fig:chiv}(color online) 
$\varepsilon_{\rm f}$ dependence of valence susceptibility $\chi_{\rm v}$ 
for $U_{\rm fc}$ from 0.0 to 1.4 at $D=1$, $V=0.5$, $n=7/8$ under $h=0$. 
Inset shows $\chi_{\rm v}$ vs. $\varepsilon_{\rm f}$ for $U_{\rm fc}=1.4$. 
} 
\end{figure}

The ground-state phase diagram at a zero magnetic field 
determined by the slave-boson mean-field theory is shown in Fig.~\ref{fig:h0PD}. 
The FOVT line represented by the solid line with open triangles in Fig.~\ref{fig:h0PD} satisfies the relation $\varepsilon_{\rm f}+U_{\rm fc}\bar{n}_{\rm c}\sim \mu$ with $\mu$ being the chemical potential~\cite{OM,WIM,M07} in the mean-field framework~\cite{noteMF}. This implies that the $f^{1}$ state with the energy $\varepsilon_{\rm f}+U_{\rm fc}\bar{n}_{\rm c}$ and the $f^{0}$ state with a conduction electron at the Fermi level with the energy $\mu$ are degenerate, giving rise to the valence transition~\cite{OM}. 
The FOVT line terminates at the QCP. 
The QCP in the $\varepsilon_{\rm f}$-$U_{\rm fc}$ plane is identified to be 
$(\varepsilon_{\rm f}^{\rm QCP}, U_{\rm fc}^{\rm QCP})=(0.356,1.464)$, 
at which the jump in $n_{\rm f}$ disappears 
and the valence susceptibility 
\begin{eqnarray}
\chi_{\rm v}\equiv -\frac{\partial^{2} \langle H \rangle}{\partial \varepsilon_{ \rm f}^{2}}
=-\frac{\partial \bar{n}_{\rm f}}{\partial \varepsilon_{\rm f}}
\label{eq:chiv}
\end{eqnarray}
diverges.
Namely, valence fluctuations diverge at the QCP. 
Even for $U_{\rm fc}<U_{\rm fc}^{\rm QCP}$, 
enhanced valence fluctuations remain~\cite{WTMF,WM}, as shown by the dashed line in Fig.~\ref{fig:h0PD}, where $\chi_{\rm v}$ has a maximum as a function of $\varepsilon_{\rm f}$ for each $U_{\rm fc}$ (see Fig.~\ref{fig:chiv}). 
The valence-crossover line with enhanced $\chi_{\rm v}$ 
regarded as a straight extension of the FOVT line to the $U_{\rm fc}<U_{\rm fc}^{\rm QCP}$ regime 
implies that the valence fluctuations are a result of the degeneracy of the $f^0$ and $f^1$ states, as mentioned above.
The characteristic energy scale of the system, the so-called Kondo temperature, 
which is defined as $T_{\rm K}\equiv \bar{\varepsilon}_{{\rm f} \sigma}-\mu$, 
is estimated to be $T_{\rm K}=0.074$ at the QCP.

Note that the large Fermi surface is realized in both the Kondo and MV states: 
Namely, the number of f electrons is always included in the total Fermi volume 
while keeping the hybridization between f and conduction electrons, 
as confirmed by the DMRG calculation~\cite{WIM}. 
This is consistent with the existence of the QCP in the ground-state phase diagram, 
since by detouring around the QCP, the Kondo and MV states can be adiabatically connected, 
with Luttinger's sum rule satisfied. 

Note that 
when the hybridization $V$ is decreased (increased), 
the QCP is shifted to a larger (smaller)  $U_{\rm fc}$ and $|\varepsilon_{\rm f}|$ position 
in Fig.~\ref{fig:h0PD}, 
as confirmed by the DMRG calculation as well as the slave-boson mean-field theory~\cite{WIM}. 
When pressure is applied to the Ce-based compounds such as $\rm CeCu_2Ge_2$ and $\rm CeIrIn_5$, 
the anion approaches the 4f electron at the Ce site, which makes the 
f-electron level $\varepsilon_{\rm f}$ increase. 
Since the interorbital Coulomb repulsion $U_{\rm fc}$ and the hybridization $V$ also increase 
under pressure, applying pressure is considered to draw a trajectory line from 
the left and bottom position to the right and top position 
in Fig.~\ref{fig:h0PD}. 
On the other hand, when pressure is applied to Ce metal, it is expected that the 
increase in the hybridization $V$ will be prominent rather than $\varepsilon_{\rm f}$, 
because of the monoelemental constitution of the metal. 
Hence, applying pressure is considered to draw a trajectory line 
from the left and bottom position to the right and top position 
in the $V$-$U_{\rm fc}$ plane, 
instead of 
in the $\varepsilon_{\rm f}$-$U_{\rm fc}$ plane 
in Fig.~\ref{fig:h0PD}. 
This is consistent with the estimation of model parameters for the $\gamma$-$\alpha$ transition 
in Ce metal based on the analysis of photoemission spectra 
using the single-impurity Anderson model~\cite{Allen}. 
The surface of FOVT exists in the parameter 
space of $\varepsilon_{\rm f}$, $U_{\rm fc}$, and $V$ 
for the ground state. A trajectory line is drawn 
in the space for the corresponding experimental parameter, such as pressure in Fig.~\ref{fig:TPH}. 

In Ce metal, the X-ray $L_{\rm III}$ edge absorption spectra 
led to the conclusion that  
the Ce valence jumps between Ce$^{+3.03}$ ($\gamma$ phase) 
and Ce$^{+3.14}$ ($\alpha$ phase) at $T=300$~K~\cite{Wohlleben}. 
One may think that the valence change seems to be too small in comparison with 
our theoretical result shown in the inset of Fig.~\ref{fig:h0PD}. 
First, we should note that the above measurement was performed at a rather high temperature 
(see Fig.~\ref{fig:Ce}(a)), and hence the magnitude of the $\bar{n}_{\rm f}$ jump at $T=0$ 
in the inset of Fig.~\ref{fig:h0PD} should be markedly reduced by thermal fluctuation effects. 
While, for qualitatively accurate comparison, it is necessary to use 
realistic band structures for the f and conduction bands, the 
momentum dependence of the hybridization, and the Coulomb interactions 
$U_{\rm fc}$ and $U$ in the model (\ref{eq:PAM}), it should be noted that 
when the symmetry of the wave function of hybridized conduction electrons is the same as that 
of f electrons, the X-ray absorption measurement detects 
the spectra 
as it comes from f electrons. 
Hence, there is a tendency that this type of measurement underestimates 
the magnitude of the valence jump. 

A key parameter for describing valence instabilities 
is the interorbital Coulomb repulsion $U_{\rm fc}$, as mentioned above. 
For Ce metal, the onsite $U_{\rm fc}$ has a considerable value 
giving $U_{\rm fc}>U_{\rm fc}^{\rm QCP}$~\cite{TM} 
and hence the FOVT is considered to occur at a very high critical end temperature $T_{\rm CEP}$ 
of about 600~K. 
In the Ce-based compounds such as $\rm CeCu_2Ge_2$, $\rm CeCu_2Si_2$, and 
$\rm CeCu_2(Ge_xSi_{1-x})_2$, 
$U_{\rm fc}$ originates from its 
intersite Coulomb repulsion and hence $U_{\rm fc}$ is reduced from the onsite value, 
which seems to comparable to $U_{\rm fc}^{\rm QCP}$. 
Namely, these compounds seem to be located in the valence crossover regime (although 
the sharp peak of the residual resistivity and the sharp drop of the $T^{2}$ coefficient 
in the resistivity under pressure suggest that these are 
rather close to the QCP~\cite{jaccard,holms,yuan,yuan2}). 

Here, we should also comment on the magnetically ordered phase, which can appear in the ground-state phase diagram in Fig.~\ref{fig:h0PD} depending on the strength of $V$. Although we focus on the nature of the FOVT line with the QCP and hence the magnetically ordered phase is not shown in Fig.~\ref{fig:h0PD}, the magnetically ordered phase is considered to be realized in the Kondo regime, which is basically located in the small-$\varepsilon_{\rm f}$ region in Fig.~\ref{fig:h0PD}. Then, in the valence-crossover regime for $U_{\rm fc}<U_{\rm fc}^{\rm QCP}$, as $\varepsilon_{\rm f}$ increases, the magnetic order is suppressed and the paramagnetic metal phase appears. In the paramagnetic metal phase, as $\varepsilon_{\rm f}$ further increases, the Kondo state is changed to the MV state at the valence-crossover point represented by the dashed line in Fig.~\ref{fig:h0PD}. 
We note that in the Kondo regime near the QCP in Fig.~\ref{fig:h0PD}, the superconducting correlation is enhanced, which was shown by the slave-boson mean field theory taking into account the Gaussian fluctuations~\cite{OM} and the DMRG calculation~\cite{WIM} applied to eq.~(\ref{eq:PAM}). This seems to correspond to the $T$-$P$ phase diagrams of CeCu$_2$Ge$_2$~\cite {jaccard}, $\rm CeCu_2Si_2$~\cite{holms}, and $\rm CeCu_2(Ge_xSi_{1-x})_2$~\cite{yuan,yuan2}, where with $P$ application, the antiferromagnetic (AF) order is suppressed and in the narrow pressure range just before a sharp valence increase of Ce, the superconducting transition temperature is enhanced. We also note that the reason why the superconducting correlation is enhanced was clarified by the unbiased calculation~\cite{WIM}: The coherence of electrons with large valence fluctuations is enhanced in the Kondo regime near the QCP, giving rise to an enhanced pairing correlation 
(see ref.~\citen{WIM} for details).

We also note that the nontrivial result has been obtained by the DMRG calculation on the model (2): 
Total charge compressibility is defined by 
\begin{eqnarray}
\kappa\equiv
\frac{1}{4n^{2}}
\frac{\partial (2n)}{\partial \mu}, 
\label{eq:comp}
\end{eqnarray}
with $2n$ being the total filling and $2n=\bar{n}_{\rm f}+\bar{n}_{\rm c}$ 
not diverging even at the QCP~\cite{WIM,WM}.
This is in sharp contrast to the mean-field result where 
the phase separation is accompanied by the FOVT in the ground-state phase diagram, 
giving rise to diverging $\kappa$. 
In the mean field framework, the valence fluctuation, i.e., relative charge fluctuation 
diverges at the valence QCP, which triggers the total charge instability as well. 
However, when quantum fluctuations and electron correlations are taken into account correctly, 
$\chi_{\rm v}$ diverges at the valence QCP, but $\kappa$ remains finite~\cite{WIM,WM}.
This has been clarified to be due to the fact that 
the order parameter of the valence transition $n_{\rm f}$ is {\it not} a conserving 
quantity; $[n_{\rm f},H]\ne 0$~\cite{WIM}. 
The system can be unstable with respect to the relative charge fluctuation 
while keeping the total charge stable (see ref.~\citen{WIM} for details). 
Hence, it is predicted that, when the material parameters 
could be experimentally tuned close to the valence QCP, the compressibility is 
\begin{eqnarray}
\kappa=-\frac{1}{V_{\rm o}}
\left(
\frac{\partial V_{\rm o}}{\partial P}
\right)_{N_{\rm e}}
\label{eq:kappa2}
\end{eqnarray}
with $N_{\rm e}$ being the total electron number 
not showing divergence at least in electronic origin. 

\begin{figure}
\includegraphics[width=80mm]{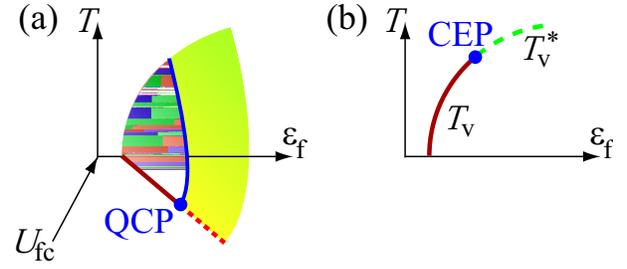}
\caption{\label{fig:PD} 
(Color online) 
(a) Schematic phase diagram in the $\varepsilon_{\rm f}$-$U_{\rm fc}$-$T$ space 
for a certain $V$ and $U(>U_{\rm fc})$. 
The FOVT surface (gray surface) 
with the critical end line (bold line) reaching on $T=0$ at the 
QCP continues to the valence-crossover surface 
(light-gray surface). 
(b) A two-dimensional cut of (a) for a certain $U_{\rm fc}>U_{\rm fc}^{\rm QCP}$. 
The FOVT line (solid line) $T_{\rm v}(\varepsilon_{\rm f})$ terminates at the CEP (filled circle), 
which continues to the valence-crossover line (dashed line) $T_{\rm v}^{*}(\varepsilon_{\rm f})$. 
}
\end{figure}

When we add a temperature axis to Fig.~\ref{fig:h0PD} at $h=0$, the phase diagram of the 
$T$-$\varepsilon_{\rm f}$-$U_{\rm fc}$ space
is shown schematically in Fig.~\ref{fig:PD}(a). 
The first-order transition surface continues to the valence crossover surface. 
The boundary between the two surfaces forms a critical end line, which reaches  
$T=0$~K, forming the QCP. 
At the critical end line as well as at the QCP, the valence susceptibility (\ref{eq:chiv}) 
diverges, i.e., 
$\chi_{\rm v}=\infty$. 
Note that, even at the valence-crossover surface, valence fluctuations develop well 
as shown in Fig.~\ref{fig:chiv}. 
Figure~\ref{fig:PD}(b) shows a two-dimensional cut of Fig.~\ref{fig:PD}(a) 
for a certain $U_{\rm fc}>U_{\rm fc}^{\rm QCP}$: The FOVT 
line (a solid line) terminates at the critical end point (a filled circle), 
which continues to the valence crossover line (a dashed line). 
The reason why $T_{\rm v}$ is an increasing function of $\varepsilon_{\rm f}$ can be understood 
from the Claudius-Clapeyron relation 
\begin{eqnarray}
\frac{d T_{\rm v}}{d \varepsilon_{\rm f}}=
\frac{n^{\rm f}_{\rm K}-n^{\rm f}_{\rm MV}}{S_{\rm K}-S_{\rm MV}}, 
\label{eq:CCR2}
\end{eqnarray}
where $n^{\rm f}$ and $S$ denote the number of f electrons (or f holes) per site 
and the entropy, respectively. 
Note here that the increase in $\varepsilon_{\rm f}$ parameterizes that of 
pressure in our model. 
We have $d T_{\rm v}/d \varepsilon_{\rm f}>0$, since 
$n^{\rm f}_{\rm K}>n^{\rm f}_{\rm MV}$ and $S_{\rm K}>S_{\rm MV}$ are satisfied 
at least in the deep first-order transition region for $U_{\rm fc}>U_{\rm fc}^{\rm QCP}$. 
Namely, to achieve the free-energy gain caused by the larger entropy, 
the Kondo phase is realized on the higher-$T$ side. 
Owing to the thermodynamic third law, the first-order transition 
temperature $T_{\rm v}$ should be perpendicular to $\varepsilon_{\rm f}$ for $T_{\rm v}\to 0$ 
as understood from eq.~(\ref{eq:CCR2}), 
which was shown in general cases in ref.~\citen{WI_thermo}.

Valence instability is considered to be coupled to volume variation. 
Hence, in our model (\ref{eq:PAM}), the effect of the hybridization 
also plays an important role, which might share common aspects with the 
Kondo-volume-collapse scenario for Ce metal~\cite{Dzero}. 
However, note that the Kondo-volume-collapse scenario assumes a special volume dependence 
on the Kondo coupling to realize the first-order transition, 
whose validity should be carefully examined. 
Actually, it has been pointed out that the Kondo-volume-collapse scenario is {\it not} 
consistent with the isostructural FOVT in $\rm YbInCu_4$ 
by Sarrao~\cite{sarraoPhysicaB1999}: 
The measured Gr{\"u}neisen parameter $\Gamma=-d{\rm ln}T_{\rm K}/d{\rm ln}V_{o}=43$ 
leads to the conclusion that 
the difference in $T_{\rm K}$ 
at the first-order transition should be $\Delta T_{\rm K}=10$~K for the measured 
volume change, $\Delta V_{o}/V_{o}=0.005$, which is much less than 
the measured one $\Delta T_{\rm K}$ $\sim 400$~K 
($T_{\rm K}$ and $V_{o}$ denote the Kondo temperature and volume, respectively). 
Thus, the volume change is too small to explain the $T_{\rm K}$ change. 
In our approach (\ref{eq:PAM}), the parameters $(U_{\rm fc}, \varepsilon_{\rm f})$ 
for each material determine whether they show the FOVT, 
or the valence crossover, when $T$ or pressure (or chemical composition) is changed. 
Our scenario does not need to assume the special volume dependence 
on Kondo coupling to 
cause an FOVT differently from that in the Kondo-volume-collapse scenario.

\subsection{Effects of magnetic field}

\subsubsection{Results by slave-boson mean-field theory}

\begin{figure}[t]
\includegraphics[width=80mm]{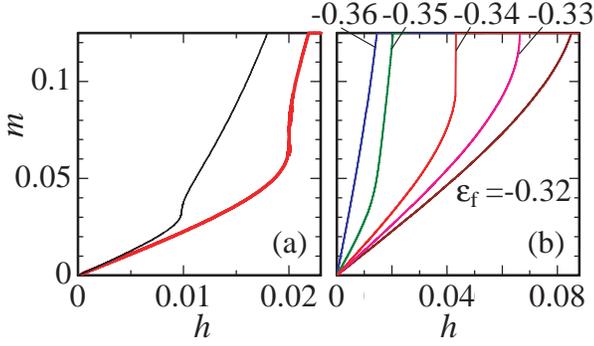} 
\caption{\label{fig:mhcurve}(color online)  
$m$-$h$ curve 
(a) for $(\varepsilon_{\rm f}, U_{\rm fc})=(-0.354,1.458)$ (thin line) 
and $(-0.349, 1.442)$ (bold line), 
and (b) for $\varepsilon_{\rm f}$ ranging from $-0.36$ to $-0.32$ 
at $U_{\rm fc}=1.42$. In both cases, $D=1$, $V=0.5$ at $n=7/8$. 
}
\end{figure}

\begin{figure}
\includegraphics[width=85mm]{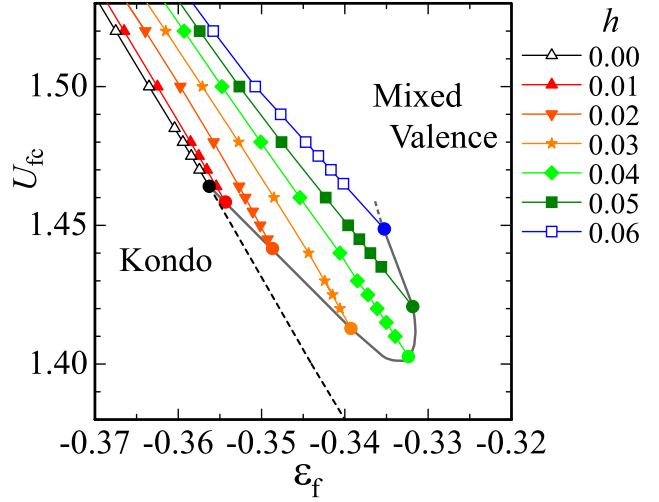}
\caption{\label{fig:hPD}(color online) Ground-state phase diagram in the plane of $U_{\rm fc}$ and 
$\varepsilon_{\rm f}$ for $D=1$ and $V=0.5$ at $n=7/8$. 
The FOVT line with a QCP for 
$h=0.00$ (open triangle), $h=0.01$ (filled triangle) $h=0.02$ (filled inverse triangle), 
$h=0.03$ (filled star), $h=0.04$ (filled diamond), 
$h=0.05$ (filled square), and $h=0.06$ (open square). 
The shaded line connects the QCP under $h$, which is shown as a visual guide. 
The dashed line represents the valence-crossover points at which $\chi_{\rm v}$ has a maximum 
as a function of $\varepsilon_{\rm f}$ for each $U_{\rm fc}$ at $h=0.00$. 
} 
\end{figure}

When we apply a magnetic field to the Hamiltonian eq.~(\ref{eq:PAM})
we find a remarkable result 
in the valence-crossover regime for $U_{\rm fc}<U_{\rm fc}^{\rm QCP}$. 
Figure~\ref{fig:mhcurve}(a) shows 
the relation of the magnetization 
$m=\sum_{i}\langle S_i^{{\rm f}z} + S_i^{{\rm c}z}\rangle/N$ vs $h$ 
for $(\varepsilon_{\rm f}, U_{\rm fc})$$=$
$(-0.354,1.458)$
(thin line) 
and 
$(-0.349, 1.442)$ 
(bold line), indicating that 
the metamagnetism (defined by the diverging magnetic susceptibility $\chi=\partial m/\partial h=\infty$) 
emerges at $h=h_{\rm m}=0.01$ and 0.02, respectively. 
To clarify its origin, we determine the FOVT line as well as the QCP 
under the magnetic field. 
The result is shown in Fig.~\ref{fig:hPD}. 
It is found that the FOVT line extends to the MV regime and the location of the QCP 
shifts to a smaller-$U_{ \rm fc}$ and smaller-$|\varepsilon_{ \rm f}|$ direction, 
when $h$ is applied. 
This low-$h$ behavior of the FOVT line agrees with the low-temperature limit of $T_{\rm v}$ 
discussed in \S~2, 
in which the FOVT line extends up to the higher pressure region as $h$ is increased 
as shown in Fig.~\ref{fig:TPH}(a). 

In Fig.~\ref{fig:mhcurve}(b)
we show the $m$-$h$ curve 
at $U_{\rm fc}=1.42$ for $\varepsilon_{\rm f}$ values ranging from $-0.32$ to $-0.36$. 
The Kondo temperature $T_{\rm K}$ at $h=0$ is estimated as 
0.0353, 
0.0873, 
0.1346, 
0.1611, 
and 
0.1823 
for $\varepsilon_{\rm f}=-0.36$, $-0.35$, $-0.34$, $-0.33$, and $-0.32$, respectively. 
From these results, the mechanism is understood as follows:
At $h=0$, $T_{\rm K}$ is originally large for $\varepsilon_{\rm f}=-0.32$ and $-0.33$, 
since the system is in the MV regime. 
However, by applying $h$, the QCP is induced, which makes reduces $T_{\rm K}$, 
since the system is forced to be closer to the Kondo regime by $h$. 
At a magnetic field $h=h_{\rm m}$ where the QCP is reached, metamagnetism occurs 
with the singularity $\delta m \sim \delta h^{1/3}$ 
as shown by Millis {\it et al}.~\cite{Millis2} (see Fig.~\ref{fig:mhcurve}(a)).   
The uniform susceptibility diverges at the QCP of the FOVT. 
The phenomena lead to the emergence of strong ferromagnetic fluctuations. 
Namely, valence fluctuations diverge there, which are 
essentially charge fluctuations. 
On the other hand, for $\varepsilon_{\rm f}=-0.35$ and $-0.36$, 
the QCP is not reached, so that no metamagnetism appears. 

Note that this mechanism is different from the ordinary metamagnetism 
emerging when the magnetic field is applied to the Kondo state,
which has been discussed as the origin of the metamagnetism observed 
in $\rm CeRu_2Si_2$~\cite{Mignot,sakakibara,Aoki,Evans,Ono,wataKLM,MI}. 
Namely, the present metamagnetism is caused by the field-induced QCP 
in the valence-crossover regime at $h=0$ (for moderate $\varepsilon_{\rm f}$ and 
not large $U_{\rm fc}<U_{\rm fc}^{\rm QCP}$ in Fig.~\ref{fig:hPD}), while the ordinary one is 
caused in the Kondo regime (for deep $\varepsilon_{\rm f}$ and hence 
$\bar{n}_{\rm f}\sim 1$, see Fig.~\ref{fig:hPD}). 
It corresponds to the collapse of antiferromagnetic correlations and 
the emergence of ferromagnetic fluctuation in a sharp $h$ window.

An interesting result is shown in 
Fig.~\ref{fig:hPD}, which exhibits a nonmonotonic $h$ dependence of the QCP:
As $h$ increases, 
the QCP shows an upturn at approximately $h=0.04$, 
which is comparable to $T_{\rm K}$ at the QCP for $h=0$. 
The upturn of the QCP  
has also been confirmed for a constant density of states $N_{0}(\varepsilon)=1/(2D)$. 
This nontrivial field dependence of the QCP appears in the valence-crossover regime 
for $U_{\rm fc}<U_{\rm fc}^{\rm QCP}$ in Fig.~\ref{fig:hPD} 
in sharp contrast to the regime for $U_{\rm fc}>U_{\rm fc}^{\rm QCP}$, 
where $T_{\rm v}$ is monotonically suppressed by $h$ as shown in Fig.~\ref{fig:TPH}(b). 

\subsubsection{RPA description of QCP}

The nonmonotonic behavior can be understood from the structure of the valence 
susceptibility $\chi_{\rm v}$, which is given essentially by the RPA, as discussed 
in ref.~\citen{M07}.  Namely, it is given as 
\begin{equation}
\chi_{\rm v}(q)\approx 
\frac{\chi_{ \rm fc}^{(0)}(q)}
{1-U_{ \rm fc}\chi_{ \rm fc}^{(0)}(q)},
\label{extra1}
\end{equation}
where $\chi_{ \rm fc}^{(0)}$ is the bubble diagram composed of f and 
conduction electrons.  In the Kondo regime ($h\lsim T_{ \rm K}$), where 
f electrons have a predominant spectral weight at approximately 
$\epsilon\sim \varepsilon_{ \rm f}$ with 
width $\Delta\simeq \pi V^{2}N(\varepsilon_{ \rm F})$, $\chi_{ \rm fc}^{(0)}$ is 
estimated as $\chi_{ \rm fc}^{(0)}\approx 1/|\varepsilon_{ \rm f}|$ and is 
shown to be an increasing function of $h$.  
Therefore, $U_{ \rm fc}^{\rm QCP}$ 
decreases as $h$ is applied until it reaches around $h\sim T_{ \rm K}$, and 
$|\varepsilon_{ \rm f}^{\rm QCP}|\approx U_{ \rm fc}^{\rm QCP}$ also 
decreases.  For $h\gsim T_{ \rm K}$, mass enhancement ($\sim 1/z$) is quickly 
suppressed and 
the MV regime 
is reached.  
Then, 
$\chi_{ \rm fc}^{(0)}$ is given as 
$\chi_{ \rm fc}^{(0)}\approx 1/\Delta <1/|\varepsilon_{ \rm f}|$ with using the help of 
shift of the f level towards the Fermi level, i.e., 
$\varepsilon_{ \rm f}\to\varepsilon_{ \rm f}+U_{ \rm fc}\delta
\bar{n}_{\rm c}$ ($\delta\bar{n}_{\rm c}$ being the 
change in the number of conduction electrons per site due to entry into  
the MV regime), 
so that $U_{ \rm fc}^{\rm QCP}$ becomes larger than 
$U_{ \rm fc}^{\rm QCP}(h\sim T_{ \rm K})$.  

\subsubsection{Distinct energy scale from Kondo temperature}

%
The magnetic field $h_{\rm m}$ at QCP when CEP collapses in a magnetic field 
corresponds to the difference in $T_{\rm K}$ between $h=0$ and $h=h_{\rm m}$: 
\begin{eqnarray}
h_{\rm m}\sim \Delta T_{\rm K}^{\rm QCP}=
T_{\rm K}^{\rm QCP}(h \ne 0)-T_{\rm K}^{\rm QCP}(h=0).
\end{eqnarray}
A new energy scale distinct from $T_{\rm K}$ reproduces the closeness 
to the valence QCP. 
Under a magnetic field, the proximity of the intermediate-valence crossover regime to QCP can 
lead to the emergence of metamagnetism with a jump of 
$m$ without initially showing the temperature-driven FOVT at $h=0$. 
Thus, it is a field-reentrant FOVT. 


\subsubsection{DMRG analysis}

\begin{figure}[t]
\includegraphics[width=80mm]{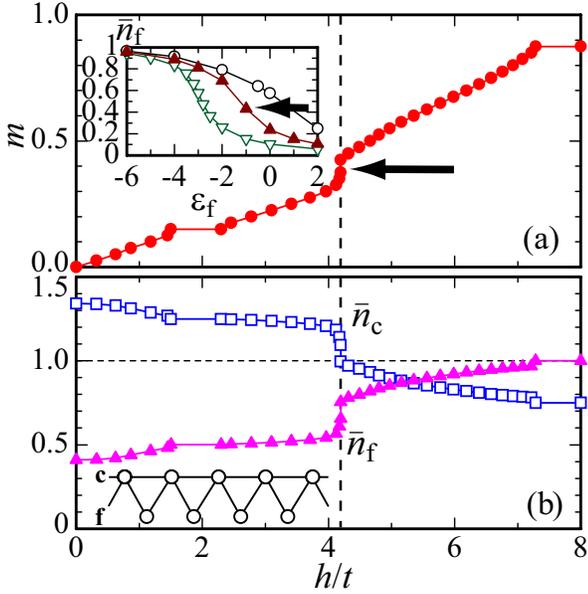}
\caption{\label{fig:DMRG}(color online) 
Magnetization process for $\varepsilon_{k}=-2\cos(k)$,
$V=1$, $U=10^{4}$, $\varepsilon_{\rm f}=-1$ and $U_{\rm fc}=1$ at $n=7/8$ 
calculated by the $T=0$ DMRG method: 
(a) $m$-$h$ curve (filled circle). 
An arrow indicates the metamagnetic transition. 
Inset: 
$\varepsilon_{\rm f}$ dependence of $\bar{n}_{\rm f}$ extrapolated to the bulk limit 
for $U_{\rm fc}=0$ (open circle), $U_{\rm fc}=1$ (filled triangle) and $U_{\rm fc}=2$ (open triangle) 
at $h=0$. 
An arrow indicates $\varepsilon_{\rm f}=-1$. 
(b) $\bar{n}_{\rm f}$ (filled triangle) and $\bar{n}_{{\rm c}}$ (open square). 
Inset: Lattice structure used in the calculation.
In (a) and (b), dashed lines represent $h=h_{\rm m}$. 
}
\end{figure}

To examine the mechanism more precisely, 
we apply the density-matrix-renormalization-group (DMRG) method~\cite{white1,white2} 
to eq.~(\ref{eq:PAM}) in one dimension.  
Since valence fluctuations are basically ascribed to be of atomic origin, 
the fundamental properties are expected to be captured even in one dimension~\cite{WIM}. 
We show here the results 
for $V=1$ and $U=10^4$  in eq.~(\ref{eq:PAM}) at $n=7/8$ 
on the lattice with $N=40$ sites (open-boundary condition), as 
illustrated in the inset of Fig.~\ref{fig:DMRG}(b). 
Here, the transfer term for conduction electrons
is expressed as 
$-\sum_{i=1,\sigma}^{N-1}(c_{i,\sigma}^{\dagger}c_{i+1,\sigma}+{\rm H. C.})$. 
This lattice may be regarded as a one-dimensional mimic of 
$\rm CeIrIn_5$ and $\rm YbXCu_4$, which will be discussed in detail in \S~4. 

For $h=0$, 
the relation of 
$\bar{n}_{\rm f}$ vs $\varepsilon_{\rm f}$ for $U_{\rm fc}=0.0$, 1.0, and 2.0 
is shown in the inset of Fig.~\ref{fig:DMRG}(a). 
As $U_{\rm fc}$ increases, the change in $\bar{n}_{\rm f}$ 
as a function of $\varepsilon_{\rm f}$ 
becomes sharp. 
We show in Fig.~\ref{fig:DMRG}(a) the magnetization 
$m=\sum_{i}\langle S_i^{{\rm f}z}+S_i^{{\rm c}z} \rangle/N$ 
in the MV state for $\bar{n}_{\rm f}$ 
indicated by an arrow in the inset of Fig.~\ref{fig:DMRG}(a), which is obtained at $h=0$. 
A plateau appears at $m=1-n=1/8$, which is expected to disappear 
if we take a 
more realistic choice of parameters, e.g., the momentum dependences of 
$V$ and $\varepsilon_{\rm f}$. 
The main result is that metamagnetism emerges, as indicated by an arrow. 

The increase in $\bar{n}_{\rm f}$ with a simultaneous decrease in $\bar{n}_{{\rm c}}$ at $h=h_{\rm m}$ 
is shown in Fig.~\ref{fig:DMRG}(b). 
It is caused by the field-induced extension of the QCP to the MV regime. 
Namely, these results indicate that 
the mean-field conclusion is not altered 
even after properly taking into account the quantum fluctuations and electron correlations. 

To further explore the nature of this metamagnetism, 
we calculate $\sum_{i}\langle S^{{\rm f}z}_i \rangle/N$ and 
$\sum_{i}\langle S^{{\rm c}z}_i \rangle/N$, 
and we find that 
$\langle S^{{\rm c}z}_i \rangle$ decreases slightly at $h=h_{\rm m}$, 
while $\langle S^{{\rm f}z}_i \rangle$ increases considerably~\cite{WTMF}. 
Since the Kondo cloud is still formed even at $h=h_{\rm m}$, i.e., 
$\langle {\bf S}^{\rm f}_{i}\cdot {\bf S}^{{\rm c}}_{i}\rangle <0$, 
the decrease of $\langle S^{{\rm c}z}_i \rangle$ is ascribed to 
{\it the field-induced Kondo effect}, which is a consequence of 
the energy benefit by both the Kondo effect and the Zeeman effect. 
Although this mechanism itself has been known to work in the Kondo regime~\cite{wataKLM}, 
this result shows that such a mechanism works in the MV regime 
as a driving force of the field-induced valence QCP.

\subsubsection{Locality of valence transition}

The DMRG calculation has shown that 
the magnetic susceptibility diverges at the QCP of the FOVT 
under a magnetic field, 
as shown in Fig.~\ref{fig:DMRG}(a). 
It is consistent with the slave-boson mean-field theory 
shown in Fig.~\ref{fig:mhcurve}(a). 
The simultaneous divergence 
of the magnetic and valence susceptibilities at the QCP 
has been confirmed by the unbiased calculation; 
in addition 
the one-dimensional calculation 
has been shown to be 
not special. 
It captures the essential physics of 
valence transition. 
The main reason is the locality of the valence transition. 
Namely, the valence transition has a local atomic origin.  
Because of this local nature, its basic properties and the ground-state phase diagram of 
the valence transition do not depend on spatial dimensions. 
Actually, the DMRG calculation in one dimension has been known to give essentially the same phase diagram 
determined by the slave boson mean field theory (see ref.~\citen{WIM}). 
Recently, 
the same phase diagram has also been obtained  by the dynamical mean field theory 
in infinite dimension~\cite{Saiga}. 

The mean-field result supported by the DMRG result implies that the RPA approach 
described in \S~3.3.2 
is qualitatively correct.
The validity of the RPA approach is also ensured 
by the perturbation renormalization group argument. 
The dynamical exponent of critical valence fluctuations is basically given by $z_{\rm d}=3$, 
and the condition $d+z_{\rm d}\ge 6$ is marginally satisfied 
for $d=3$-dimensional systems~\cite{M07}. 
This is a condition for the third-order term in the free-energy expansion to be irrelevant, 
as discussed by Hertz for the fourth-order term~\cite{Hertz}. 
Then, the universality class of the valence QCP essentially belongs 
to the Gaussian fixed point, which justifies the RPA approach.

\subsubsection{Temperature dependences of FOVT and valence crossover 
under magnetic field}

As shown in \S~2, the FOVT temperature $T_{\rm v}(h)$ 
is suppressed by applying $h$. 
We note, however, that the temperature dependence of $T_{\rm v}(h)$ can change 
according to the location in the ground-state phase diagram (see Fig.~\ref{fig:hPD}). 
When the system is located at the deep first-order transition side, i.e., 
at $U_{\rm fc}\gg U_{\rm fc}^{\rm QCP}$ 
in Fig.~\ref{fig:hPD}, $T_{\rm v}(h)$ is a decreasing function of $h$, 
as shown in Fig.~\ref{fig:TPH}(b). 
Actually, the proof by the Claudius-Clapeyron relation is based on the fact that 
$m_{\rm K}>m_{\rm MV}$ and $S_{\rm K}>S_{\rm MV}$ at $T_{\rm v}(h)$ in eq.~(\ref{eq:CCR}); 
this is justified at the deep first-order transition side. 
Ce metal and YbInCu$_4$ correspond to this case. 

On the other hand, near the QCP ($U_{\rm fc}\sim U_{\rm fc}^{\rm QCP}$) 
as well as in the valence-crossover regime at $U_{\rm fc}< U_{\rm fc}^{\rm QCP}$, 
a different situation can arise: 
In the case that $\bar{n}_{\rm f}$ at the QCP is not very close to 1 
(as is certainly the case for Yb systems~\cite{Knebel}), the above relations on 
magnetization and entropy can change. 
In addition, details of the trajectory line of the QCP under $h$ may 
severely affect the $h$ dependence of $T_{\rm v}(h)$. 
For example, CeIrIn$_5$, which will be discussed in detail in the \S~4, 
is considered to be located 
in the valence-crossover regime for $U_{\rm fc}< U_{\rm fc}^{\rm QCP}$ and 
the field-induced $T_{\rm v}(h)$ is considered to increase under $h$, 
in contrast to those in the cases of Ce metal and YbInCu$_4$. 

We have already stressed that a valence-crossover surface 
exists in the $T$-$\varepsilon_{\rm f}$-$U_{\rm fc}$ space as shown in Fig.~\ref{fig:PD}(a). 
Hence, even for $U_{\rm fc}\ll U_{\rm fc}^{\rm QCP}$, 
the valence crossover surface will be induced by applying a magnetic field, 
giving rise to an increase in 
magnetization. 
The enhancement of the magnetic susceptibility 
at a certain magnetic field (see Figs.~\ref{fig:hPD} and \ref{fig:PD}(a)) 
will lead to a pseudo-metamagnetic effect. 
Thus, the valence-crossover temperature $T_{\rm v}^{*}(h)$ also forms a line 
in the $T$-$h$ phase diagram.

When $T_{\rm v}^{*}(h)$ is induced by applying $h$, 
if the system is close to the QCP (namely, is located closely to the 
filled circles in Fig.~\ref{fig:hPD} under $h \ge 0$), the following are expected to be 
observed in the $T$-$h$ phase diagram: 
(a) The $T$-linear resistivity appears in the wide-$T$ range when $h$ approaches 
$h_{\rm v}^*$ at which $T_{\rm v}^*(h)$ becomes zero. 
(b) Residual resistivity is enhanced toward $h_{\rm v}^*$ 
and has a maximum at $h_{\rm v}^*$. 
(c) Magnetic susceptibility has a peak at $T=T_{\rm v}^*(h)$. 
(d) NQR frequency changes sharply at $T=T_{\rm v}^*(h)$, since the charge distribution 
at the Ce or Yb site and its surrounding ions changes owing to 
the valence change of Ce or Yb, leading to the change in their electric-field 
gradient. 
(e) The lattice constant shows a sharp change at $T_{\rm v}^*(h)$ 
and hence the magnetostriction changes sharply. 

When the FOVT $T_{\rm v}(h)$ is induced by $h$, 
the above physical quantities show discontinuous jumps at $T=T_{\rm v}(h)$. 
If the system is close to the QCP, valence-fluctuation-induced anomalies such as 
the $T$-linear resistivity will also be observed even in this case. 
Let us conduct a test with experiments. 

In a series of Yb- and Ce-based compounds, the above predictions have been actually observed, 
which will be discussed in detail in the next section.

\section{Explanation for $\bf YbXCu_4$ and $\bf CeYIn_5$}
We here discuss the potentiality of our theory to resolve outstanding puzzles observed 
in Yb- and Ce-based systems. 
First, we show how our results explain 
the isostructural FOVT observed in YbInCu$_4$ and 
the sharp contrast between YbAgCu$_4$ and YbCdCu$_4$ 
in their magnetic responses. 
Second, we focus on the peculiar magnetic response 
in CeIrIn$_5$, where the first-order transition line emerges 
in the temperature-magnetic-field phase diagram, 
giving rise to non-Fermi liquid behavior. 
Third, we argue that 
the first-order like  disappearance of antiferromagnetism (AF) and the change of 
de Haas-van Alphen (dHvA) signal observed in $\rm CeRhIn_5$ at $P\sim 2.4$~GPa 
under a magnetic field $h>10$~Tesla may be explained by our model.

\subsection{YbXCu$_4$ systems}

\subsubsection{Isostructural FOVT in YbInCu$_4$}

YbInCu$_4$ is known as a typical Yb compound that exhibits the isostructural FOVT at $T=42$~K~\cite{Felner,Kojima1989,Sarrao1996} between the high-temperature phase with Yb$^{+2.97}$ and the low-temperature phase with Yb$^{+2.84}$~\cite{Cornelious,Dallera,Matsuda2}. Namely, in the hole picture, $\bar{n}_{\rm f}$ jumps from 0.97 to 0.84 as temperature decreases. This can be understood qualitatively from the result shown in Fig.~\ref{fig:PD}(a): in the FOVT region, the larger hole-density phase is realized in the high-$T$ phase (the Kondo phase), because of the free-energy gain due to the larger entropy. Since this high-$T$ phase has a smaller f-electron number, the volume of the system is considered to be small in comparison with that for the low-$T$ phase. Hence, as temperature decreases, the first-order transition to the smaller hole-density phase (the MV phase) 
is realized with volume expansion.
The collapse of $T_{\rm v}$ under a magnetic field was found in macroscopic magnetization measurement, 
explained in phenomenological approach and well confirmed by microscopic X-ray 
experiments~\cite{Matsuda2}.

As mentioned below eq.~(\ref{eq:PAM}), the band-structure calculations as well as as photoemission measurements suggest the importance of the $V$ and $U_{\rm fc}$ terms in eq.~(\ref{eq:PAM}) at the FOVT in YbInCu$_4$. Here, we should also note the possibility that band structures such as semimetallic structures also play an important role in the FOVT as pointed out in refs.~\citen{sarraoPhysicaB1999} and \citen{Figueroa}. Although the accurate estimation of the values of the model parameters of a model Hamiltonian on the basis of first-principles calculations is an important task in the future, we here discuss the basic properties of YbXCu$_4$ on the basis of eq.~(\ref{eq:PAM}).

\subsubsection{Sharp contrast between YbAgCu$_4$ and YbCdCu$_4$}

When X=In is used to replace the other elements, YbXCu$_4$ does not 
show the FOVT, 
but shows merely the valence crossover. 
For example, X=Ag~\cite{Rossel,Severing,Sarrao} 
and X=Cd~\cite{Hiraoka,Sarrao} for YbXCu$_4$ show neither the FOVT 
nor the magnetic transition and they have the paramagnetic-metal ground state. 
The Kondo temperatures of both materials estimated from the magnetic susceptibility data~\cite{Sarrao} are nearly the same, 
$T_{\rm K}\sim 200$~K. 
A striking point is that when the magnetic field is applied to these systems, 
only X=Ag shows a metamagnetic behavior in the magnetization curve, while X=Cd merely shows the gradual increase in magnetization~\cite{Sarrao}. 

\begin{figure}
\includegraphics[width=75mm]{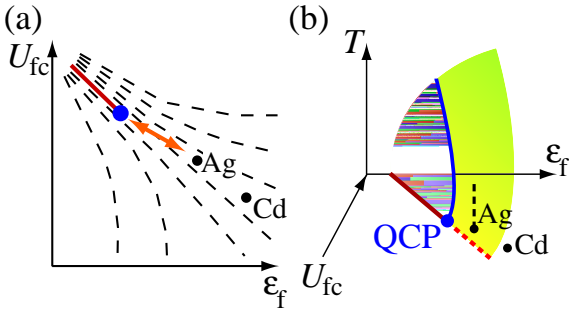}
\caption{\label{fig:ContourTk} 
(Color online) 
(a) Schematic contour plot of $\bar{n}_{\rm f}$, i.e., the Kondo temperature $T_{\rm K}$ in the $U_{\rm fc}$-$\varepsilon_{\rm f}$ plane for the Hamiltonian (1) at $h=0$. The first-order valence-transition line (solid line) terminates at the QCP (filled circle). The arrow represents a distance of $\sim 40$~T between the QCP and YbAgCu$_4$. YbCdCu$_4$ is located far away from the QCP.  
(b) Schematic $T$-$\varepsilon_{\rm f}$-$U_{\rm fc}$ phase diagram. YbAgCu$_4$ reaches 
the valence-crossover surface with a distance of about $T=40$~K. 
YbCdCu$_4$ is too far from the QCP and hence has too large a distance from the valence-crossover surface. 
}
\end{figure}

Our results explain why such a sharp increase in magnetization emerges only for X=Ag, but not for X=Cd in spite of the fact that both have nearly the same $T_{\rm K}$'s. 
Figure~\ref{fig:ContourTk} shows the schematic contour plot of the f-hole number per site, 
$\bar{n}_{\rm f}$, which can also be regarded as the contour plot of the Kondo temperature 
$T_{\rm K}$ in the $U_{\rm fc}$-$\varepsilon_{\rm f}$ plane, 
because $T_{\rm K}$ is a function of $\bar{n}_{\rm f}$, as in 
$T_{\rm K}\propto(1-\bar{n}_{\rm f})/(1-\bar{n}_{\rm f}/2)$~\cite{Rice}. 
In the small-$U_{\rm fc}$ and small-$\varepsilon_{\rm f}$ region, $\bar{n}_{\rm f}$ approaches 1, 
so that $T_{\rm K}$ becomes small. In the large-$U_{\rm fc}$ and large-$\varepsilon_{\rm f}$ region, 
$\bar{n}_{\rm f}$ is smaller than 1, giving rise to a large $T_{\rm K}$. 
As $(U_{\rm fc}, \varepsilon_{\rm f})$ approaches the QCP 
from the valence-crossover regime for $U<U_{\rm fc}^{\rm QCP}$, 
contour lines of $T_{\rm K}$'s get close, 
and at the FOVT line $T_{\rm K}$'s show discontinuous jumps.
Since the compounds with 
X=Ag and X=Cd have nearly the same $T_{\rm K}$, both are considered to be located 
near the same contour area (see Fig.~\ref{fig:ContourTk}). 
However, the compound with 
X=Ag appears located more closely to the QCP of the FOVT than X=Cd, 
with 
a valence fluctuation energy, i.e., equivalent to a Zeeman energy of approximately 40~T.
A sharp increase in magnetization will appear in the case of 
X=Ag; by applying $h\sim 40$~T, 
the field-induced QCP of the valence transition (or sharp valence crossover) 
is reached.

A sharp contrast between YbAgCu$_4$ and YbCdCu$_4$ was also observed in the $T$ dependence of 
the magnetic susceptibility $\chi(T)$~\cite{Sarrao}.  
Although both show nearly the same $\chi(0)$'s reflecting the fact that both have nearly the same 
$T_{\rm K}$'s, a broad maximum in $\chi(T)$ appears at approximately $T=40$~K only in YbAgCu$_4$, 
but a monotonic decrease in $\chi(T)$ appears in YbCdCu$_4$ as $T$ increases. 
This can be understood if YbAgCu$_4$ reaches the valence crossover surface from a distance 
of approximately $T=40$~K, while YbCdCu$_4$ reaches it at too large a temperature $T$ interval, 
as shown in Fig.~\ref{fig:ContourTk}(b). 
Since the magnetic susceptibility $\chi(T)$ has a peak at the valence crossover temperature 
$T_{\rm v}^{*}$ (see Figs.~\ref{fig:mhcurve}(a) and \ref{fig:DMRG}(a))~\cite{WM}, the peak of 
$\chi(T)$ at $T=40$~K in YbAgCu$_4$ is explained naturally. 
Indeed the proximity of $\rm YbAgCu_4$ to the QCP is reflected in the thermal volume 
expansion directly linked to the pressure dependence of the entropy: 
the volume expansion was observed below $T=40$~K~\cite{Koyama} 
simultaneously with the sharp valence crossover from Yb$^{+2.89}$ to Yb$^{+2.87}$ 
in YbAgCu$_4$, in contrast to YbCdCu$_4$~\cite{Sarrao}. 

Hence, the viewpoint of the closeness to the QCP of the FOVT 
expressed in Figs.~\ref{fig:ContourTk}(a) and \ref{fig:ContourTk}(b) not only explains the metamagnetic behavior 
but also the peak of the uniform susceptibility $\chi(T)$ consistently. 
Both phenomena are coupled with the local origin of each phenomenon.

\subsection{CeIrIn$_5$}

Our results also explain the peculiar magnetic response in 
$\rm CeIrIn_5$, which shows a jump in the $m$-$h$ curve at 42~T~\cite{Takeuchi,Kim,Parm,Capan}. 
Capan {\it et al}. have observed that residual resistivity increases, and 
the Sommerfeld constant in the specific heat shows a diverging increase toward 
the metamagnetic-transition field $\sim 25$~T. 
Furthermore, they have found that as $h$ increases, 
the power of the resistivity $\rho \sim T^{\alpha}$ at low temperatures 
decreases from $\alpha=1.5$, 
and seems to approach $\alpha=1.0$ judging 
from the fact that the convex curve appears in the $\rho\propto T^{1.5}$ plot~\cite{Capan}. 
Experimental effort should be made to properly confirm the expectation that the 
$T$-linear resistivity and the peak of the residual resistivity will be observed around $h\sim 25$~T. 
Capan {\it et al}. have pointed out that these anomalous behaviors may 
be related to the metamagnetic transition that forms a first-order-transition line 
in the $T$-$h$ phase diagram and that this may be the origin of 
the non Fermi-liquid normal state 
observed at $h=0$, although its mechanism has not yet been clarified.

Our results suggest that the mechanism is the valence fluctuation:
This can be readily understood 
if $\rm CeIrIn_5$ is located inside the enclosed area of the QCP line 
for $h\ne 0$ in Fig.~\ref{fig:hPD}. 
Namely, at $h=0$, the system is considered to be located in the gradual valence-crossover regime 
(i.e., for $U<U_{\rm fc}^{\rm QCP}$ in Fig.~\ref{fig:hPD}), since no evidence of the first-order transition has been observed at any physical quantities as a function of $T$ at $h=0$. However, when $h$ is applied, the QCP of the FOVT reaches and eventually goes across the point of the system, causing metamagnetic transition in the magnetization curve. 
Since it has been shown theoretically that the residual resistivity is enhanced near the QCP~\cite{MM} and 
that the $T$-linear resistivity is expected in a wide-$T$ region~\cite{holms}, the observed non-Fermi-liquid behavior is quite consistent.
Furthermore, the first-order transition emerges in the $T$-$h$ phase diagram 
in agreement with our predictions.

We here remind the readers of the fact that CeIrIn$_5$ and CeCoIn$_5$ have nearly the same 
crystalline-electric-field (CEF) structures~\cite{CEF} and that a change in CEF level under 
a magnetic field cannot explain the metamagnetic increase in magnetization in CeIrIn$_5$, 
as pointed out in ref.~\citen{Takeuchi}. 
Also note that almost the same Fermi surfaces in both systems have been obtained by the de Haas-van Alphen measurements as well as the first-principles band structure calculations~\cite{Harima,Shishido}. 
However, neither the enhancement of residual resistivity nor the metamagnetic-transition line in the $T$-$h$ phase diagram has been observed in CeCoIn$_5$, in contrast to that in the case of CeIrIn$_5$ at zero pressure. 
These results reemphasize 
that a distinct energy scale other than the Kondo temperature is indispensable for understanding the Ce115 systems. 

In order to directly verify the above scenario, 
detection of the Ce valence change at $T_{\rm v}(h)$ in the $T$-$h$ phase diagram 
is highly desired by measurements such as the X-ray adsorption spectra and the NQR electric gradient. 

The notable result is that $h$ scan can lead to a $h$ reentry into the valence critical domain. 
Our approach assumes a paramagnetic ground state; in some $(U_{\rm fc}, \varepsilon_{\rm f})$ 
windows, long-range magnetism will appear as in the case of YbInCu$_4$ under pressure ($P>2$~GPa). 
As suggested recently, $\rm YbRh_2Si_2$ may be a singular spectacular case where 
in the $(h, T)$ phase diagram the magnetism and valence 
fully interact~\cite{Knebel}. Such interplay also occurs in CeRhIn$_5$. 
At zero pressure, CeRhIn$_5$ is a heavy-fermion antiferromagnet with the Neel temperature $T_{\rm N}\approx 3.5$~K. At zero magnetic field, at $P>2$~GPa, a pure superconductivity phase is detected without antiferromagnetism, while under a magnetic field reentrant antiferromagnetism is detected up to $P\approx 2.4$~GPa.

\subsection{CeRhIn$_5$}

Sharp valence crossover has also been suggested in $\rm CeRhIn_5$ under pressure 
near $P=2.4$~GPa, 
where the resistivity $\rho(T=2.25~{\rm K}\approx T_{\rm SC})$, 
with $T_{\rm SC}$ being the superconducting transition temperature 
showing a peak, as well as the $T$-linear resistivity 
in the wide-$T$ range emerges~\cite{Hegger,Muramatsu,Knevel2008}. 
Hence, the sharp valence crossover originating from the QCP of the valence transition 
may be induced by applying pressure to $\rm CeRhIn_5$. 
Namely, the viewpoint of the closeness to the QCP of the FOVT is important in elucidating the $T$-$h$-$P$ (chemical doping) phase diagram of these compounds in a consistent way. 
The sharp peak of $\rho$($T=2.25$~K) cannot 
be explained by the many-body correction due to critical AF fluctuations~\cite{MN}, 
but can be understood by the enhanced valence fluctuations~\cite{MM}.  

Other lines of evidence for the crucial roles of FOVT in CeRhIn$_5$ under the magnetic 
field are as follows: 

1) According to ref.~\citen{Knevel2008}, near $P=2.4$~GPa under the magnetic 
field of 15~Tesla, both a effective mass of electrons $m^{*}/m_0$ and 
the coefficient of the
$T^{2}$ term of the resistivity $A$ exhibit a rather sharp enhancement.  However, 
$m^{*}/m_0$ scales with $A^{1/2}$, suggesting that the mass enhancement 
is mainly driven by the ``local correlation effect" (but not due to critical antiferromagnetic 
fluctuations) just as in the case of the metamagnetic transition of CeRu$_2$Si$_2$ 
discussed in refs.~\citen{wataKLM} and \citen{MI}.  
The enhancement of $m^{*}/m_0$ can be interpreted as that of the 
quasiparticle density of states near the hybridization gap or pseudo gap, 
which can be approached at approximately $P=P_{\rm v}\simeq 2.4$~GPa under the magnetic 
field as the valence changes rapidly or discontinuously with pressure.  
In other words, the first-order-like disappearance of 
antiferromagnetism~\cite{Phillips,Park2006,Knevel2008} 
and the change of the dHvA signal observed in CeRhIn$_5$ at $P=2.4$~GPa under a magnetic 
field larger than 10 Tesla\cite{Shishido2} can be naturally 
understood as a FOVT induced by the magnetic field.  

2) According to ref.~\citen{Knevel2008}, the upper critical field $H_{{\rm c}2}$ 
exhibits a rather sharp peak at $P=P_{\rm v}$ where the Fermi surface 
exhibits sharp change from ``localized" to ``itinerant" under a 
magnetic field~\cite{Shishido2}, while the superconducting transition 
temperature $T_{\rm sc}$ is essentially flat around $P=P_{\rm v}$.  This 
fact can be interpreted as an effect of the growth of the paring interaction 
due to the effect of approaching the magnetic-field-induced critical 
point of valence transition.  Such behavior reminds us of the case of 
UGe$_2$ at $P_{\rm x}=1.3$~GPa in which a magnetic field induces the 
metamagnetic transition between two ferromagnetic phases, 
leading to a sharp increase in $H_{{\rm c}2}$ with a concave 
shape~\cite{Sheikin, WatanabeUGe2}.  

3) The $P$ dependence of the low-temperature resistivity $\rho(T=2.25~{\rm K})$ has a peak 
at $P=2.4$~GPa and the emergence of the $T$-linear dependence of $\rho(T)$ 
in the vicinity of $P=2.4$~GPa~\cite{Hegger,Muramatsu,Knevel2008} 
can be naturally explained by the present mechanism~\cite{MM, holms}. 

We stress here that the ``localized"-to-``itinerant" change in electron character 
reported in the dHvA measurement~\cite{Shishido2} can be explained by the Ce-valence jump 
or sharp crossover at $P=2.4$~GPa where the number of f electrons is always included in the total Fermi volume~\cite{WIM,MI}, i.e., c-f hybridization is always switched on 
in sharp contrast to that in the Kondo breakdown scenario~\cite{Park2009}. 
Our mechanism is also consistent with the experimental fact that the effective mass of electrons is enhanced even at $P=0$ with the Sommerfeld constant $\gamma\approx 56$~mJmol$^{-1}$K$^{-2}$~\cite{Hegger}, which is about 10 times enhanced from the LaRhIn$_5$ value~\cite{Shishido,Phillips}, strongly indicating the AF state with the c-f hybridization. Furthermore, the mass enhancement observed toward $P=2.4$~GPa~\cite{Shishido2,Knevel2008} inside the AF phase can also be explained by the present mechanism. Hence, it should be stressed that the valence QCP itself is the source of locality emerging in CeRhIn$_5$ without invoking a collapse of Kondo temperature.

As shown by the phase diagram of $\rm CeRh_{x}Ir_{1-x}In_5$~\cite{kawasaki}, 
$\rm CeIrIn_5$ at ambient pressure is moderately far from the AF QCP with a distance of 
about $x\sim 0.5$. 
A slight increase in the nuclear spin-lattice relaxation rate $1/T_1T$ 
indicates that moderate spin fluctuations may exist at least at ambient pressure~\cite{kawasaki}. 
It has been also reported that 
the magnetotransport measurements under pressures can be understood from 
the effects of AF spin fluctuations~\cite{nakajima}. 
However, the clear difference between $\rm CeIrIn_5$ and $\rm CeCoIn_5$ 
emerging in the $T$-$h$ phase diagram mentioned in \S~4.2 
cannot be explained only from the sole viewpoint of the closeness to the AF QCP. 
In addition to the AF QCP, the influence of the QCP of the FOVT is 
indispensable for the comprehensive understanding.

Our present viewpoint also gives us a key to resolving the outstanding puzzle about 
the origin of the superconductivity of CeIrIn$_5$, 
whose transition temperature 
increases even though AF spin fluctuation is suppressed 
under pressure~\cite{kawasaki}.  
Since superconductivity will be enhanced near the valence QCP~\cite{OM,WIM}, 
the present viewpoint offers a new scenario that the proximity of QCP of the FOVT is 
the main origin of the superconductivity. 
The superconducting window reveals phenomenon other than spin fluctuation; the occurrence 
of superconductivity is a unique opportunity for scanning through different pairing channels. 
We have already pointed out in the introduction that in many heavy fermion compounds 
even for the magnetic QCP the interplay between spin and valence fluctuations 
is the main reason for collapse of the long-range magnetism. 

\subsection{Brief summary}
Detailed discussions for each material have been given in from \S~4.1 to \S~4.3.  They are briefly summarized as follows:

1)	The field dependences of the FOVT in Ce metal and YbInCu$_4$ are clearly explained by our mechanism. The effects of the semimatallic band structure on the FOVT as well as the qualitative evaluation of $U_{\rm fc}$ are issues to be studied in the future for the complete understanding of the valence transition of YbInCu$_4$. 

2) The metamagnetism at $h=40$~T and the peak structure in $\chi(T)$ at $T=40$~K in YbAgCu$_4$ but not in  YbCdCu$_4$ are naturally explained by our mechanism. The experimental fact that the valence crossover occurs at $T=40$~K for $h=0$ in YbAgCu$_4$ is consistent with our theory. The direct observation of the Yb-valence change under a magnetic field at approximately $h=40$~T at low temperatures is highly desired to confirm our theoretical proposal.

3)	We point out that the field-induced FOVT explains the $T$-$h$ phase diagram as well as the non-Fermi-liquid critical behavior observed in CeIrIn$_5$. It has been reported that magnetotransport measurement under pressure can be explained by AF spin fluctuations~\cite{nakajima}. We think that, in addition to the influence of the AF QCP, the viewpoint of the closeness to the QCP of the FOVT is necessary for the comprehensive understanding of CeIrIn$_5$. To examine our theoretical proposal, it is highly desired to experimentally determine whether the change in Ce valence occurs at the FOVT $T_{\rm v}(h)$ in the $h$-$T$ phase diagram. 

4)	We point out that the anomalous behaviors at approximately $P\sim 2.4$~GPa observed in CeRhIn$_5$ can be naturally explained if the FOVT or sharp valence crossover of Ce takes place at such a pressure. We think that such behavior cannot be explained solely by the AF QCP scenario. 
Our picture gives a natural explanation of the origin of the locality 
as well as the non-Fermi liquid behavior without relying on 
artificial assumptions such as the Kondo breakdown. 
It is highly desired to experimentally determine whether the Ce valence changes at approximately 
$P\sim 2.4$~GPa.

On points 3) and 4), a more quantitative evaluation of model 
parameters including CEF parameters~\cite{maehira}  is necessary for elucidating the $P$-$h$-$T$ phase diagram 
toward a complete understanding of the Ce115 system.


\section{Conclusions}

We have clarified the mechanism of novel phenomena in heavy-fermion systems 
emerging under a magnetic field 
and have discussed the significance of the proximity to the FOVT 
as a potential origin of the anomalous electronic properties of Ce- and Yb-based heavy-fermions. 
We have shown that FOVT temperature is suppressed by applying a magnetic field, 
which correctly connects the high-temperature result derived from the atomic picture of 
the valence-fluctuating ion to 
the zero-temperature limit consistently with the observations in Ce metal and $\rm YbInCu_4$. 
The important result is that 
even in intermediate-valence materials, by applying a magnetic field, 
the QCP of the FOVT is induced. 
The QCP shows a nonmonotonic field dependence in the ground-state phase diagram, 
giving rise to the emergence of metamagnetism with diverging magnetic susceptibility. 
The driving force of the field-induced QCP is clarified to be a cooperative mechanism of 
the Zeeman effect and the Kondo effect, which creates a distinct energy scale 
from the Kondo temperature. 

The use of an extended periodic Anderson model explains how quite similar valences 
may lead to quite different $h$ responses. 
Our model clarifies 
why metamagnetic behavior appears in $\rm YbAgCu_4$ but not in $\rm YbCdCu_4$, 
in spite of the fact that both have nearly the same Kondo temperatures. 
The closeness to the QCP of the FOVT gives the distinct energy scale, 
which is a key concept to understanding the properties of 
$\rm YbXCu_4$ (X=In, Ag, and Cd) systematically. 
This viewpoint also explains peculiar magnetic response in $\rm CeIrIn_5$ 
where the first-order transition line in the $T$-$h$ phase diagram appears 
with field-induced critical phenomena. 
The viewpoint of the closeness to the QCP of the FOVT 
is also indispensable for understanding $\rm CeYIn_5$ (Y=In, Co, and Rh) systematically. 

As shown in the present study, the QCP of the FOVT and its fluctuations 
exerts profound influences on Ce- and Yb-based materials as a potential origin of 
anomalous behavior. 
Most of such materials are considered to be located in the intermediate valence regime, i.e., 
in the region for $U_{\rm fc}<U_{\rm fc}^{\rm QCP}$ in Fig.~\ref{fig:hPD} due to the intersite origin of $U_{\rm fc}$. 
However, by applying a magnetic field, the valence-crossover surface as well as the critical point 
is induced, which causes various anomalies described in this paper. 
The $(P, h)$ valence transition 
mechanism clarified in this paper can be a key origin of unresolved phenomena 
in the family of the materials.


\section*{Acknowledgments}
S. W. and K. M. thank S.~Wada and A.~Yamamoto 
for showing us their experimental data prior to publication, 
with enlightening discussions on their analyses. 
They also acknowledge H.~Harima for helpful discussions about 
the band structures of Ce- and Yb-based heavy-fermions as well as their 
model parameters. 
S. W. is grateful to T.~Miyake for estimating the magnitude of the Coulomb repulsions 
in the model for Ce metal based on first-principles calculations. 
This work is supported by a Grant-in-Aid for Scientific
Research on Priority Areas (No. 18740191) from the Ministry of Education, Culture, 
Sports, Science, and Technology, Japan, 
and is supported in part by a Grant-in-Aid for Scientific Research (No. 19340099) by 
the Japan Society for the Promotion of Science (JSPS). 
J. F. is supported by the Global COE program (G10) of JSPS for supporting his 
visit of the Graduate School of Engineering Science at Osaka University where the final stage of this work was performed.
Part of our computation has been performed at the supercomputer center 
at the Institute for Solid State Physics, the University of Tokyo.



\begin{thebibliography}{99} 
\bibitem{Moriya} T. Moriya: {\it Spin Fluctuations in Itinerant Electron Magnetism} 
(Springer-Verlag, Berlin, 1985). 
\bibitem{Hertz} J. A. Hertz: Phys. Rev. B {\bf 14} (1976) 1165. 
\bibitem{Millis} A. J. Millis: Phys. Rev. B {\bf 48} (1993) 7183. 
\bibitem{Stewart} G. R. Stewart: Rev. Mod. Phys. {\bf 73} (2001) 797. 
\bibitem{MNO} K. Miyake, O. Narikiyo, and Y. Onishi: Physica B {\bf 259-261} (1999) 676. 
\bibitem{M07} K. Miyake: J. Phys.: Condens. Matter {\bf 19} (2007)  125201.
\bibitem{Bridgeman} P. W. Bridgeman: Proc. Am. Acad. Sci.{\bf 81} (1952) 165. 
\bibitem{Cevalence} K. A. Gschneidner and L. Eyring: 
{\it Handbook on the Physics and Chemistry of Rare Earths} (North-Holland, Amsterdam, 1978). 
\bibitem{Flouquet2006} J. Flouquet: in {\it Progress in Low Temperature Physics}, ed. W.~Halperin 
(Elsevier, Amsterdam, 2005) Vol. 15, p. 139. 
\bibitem{Felner} I. Felner and I. Nowik: Phys. Rev. B {\bf 33} (1986) 617. 
\bibitem{Kojima1989} K. Kojima, H. Hayashi, A.~Minami, Y.~Kasamatsu, and T.~Hihara: J. Mag. Mag. Mat. {\bf 81} (1989) 267.
\bibitem{Sarrao1996PRB} J. L. Sarrao, C. D. Immer, C. L. Benton, Z. Fisk, J. M. Lawrence, D. Mandrus, and J. D. Thompson: Phys. Rev. B {\bf 54} (1996) 12207. 
\bibitem{Mito2003} T. Mito, T. Koyama, M.~Shimoide, S.~Wada, T.~Muramatsu, T.~C.~Kobayashi, 
and J.~L.~Sarrao: Phys. Rev. B {\bf 67} (2003) 224409. 
\bibitem{Park2006PRL} T. Park, V. A. Sidorov, J. L. Sarrao, and J. D. Thompson: 
Phys. Rev. Lett. {\bf 96} (2006) 046405.
\bibitem{jaccard} D. Jaccard, H. Wilhelm, K. Alami-Yadri, and E. Vargoz: Physica B {\bf 259-261} (1999) 1. 
\bibitem{holms} A.~T. Holmes, , D. Jaccard, and K. Miyake: Phys. Rev. B {\bf 69} (2004) 024508.
\bibitem{fujiwara} K. Fujiwara, Y. Hata, K. Kobayashi, K. Miyoshi, J. Takeuchi, Y. Shimaoka, 
H. Kotegawa, T. C. Kobayashi C. Geibel, and F. Steglich: J. Phys. Soc. Jpn. {\bf 77} (2008) 123711. 
\bibitem{yuan} H.~Q. Yuan, F.~M.~Grosche, M.~Deppe, C.~Geibel, G.~Sparn, and F. Steglich: Science {\bf 302} (2003) 2104. 
\bibitem{yuan2} H.~Q. Yuan, F.~M.~Grosche, M.~Deppe, C.~Geibel, G.~Sparn, and F. Steglich: 
Phys. Rev. Lett. {\bf 96} (2006) 047008. 
\bibitem{MM} K. Miyake and H. Maebashi: J. Phys. Soc. Jpn. {\bf 71} (2002) 1007. 
\bibitem{OM} Y. Onishi and K. Miyake: J. Phys. Soc. Jpn. {\bf 69} (2000) 3955.
\bibitem{ML} P. Monthoux and G. G. Lonzarich: Phys. Rev. B {\bf 69} (2004)  064517. 
\bibitem{WIM} S. Watanabe, M. Imada, and K. Miyake: J. Phys. Soc. Jpn. {\bf 75} (2006) 043710. 
\bibitem{WTMF} S. Watanabe, A. Tsuruta, K. Miyake, and J. Flouquet: Phys. Rev. Lett. {\bf 100} (2008) 236401. 
\bibitem{defMV} We here use the terminology ``mixed valence'' 
to indicate spatially uniform and quantum-mechanically valence-fluctuating state 
with $\bar{n}_{\rm f}<1$. 
We refer to the state with a larger $\bar{n}_{\rm f}$ than 
the ``mixed-valence" state in the first-order transition as the Kondo state. 
Note that in the intermediate-coupling regime, $\bar{n}_{\rm f}$ 
in the Kondo state is smaller than 1. 
\bibitem{magCe} F. Drymiotis, J. Singleton, N. Harrison, J.~C.~Lashley, 
A.~Bangura, C.~H.~Mielke, L.~Balicas, Z.~Fisk, A.~Migliori, and J.~L.~Smith: 
J. Phys.: Condens. Matter {\bf 17} (2005)  L77-L83. 
\bibitem{Immer} C. D. Immer, J. L. Sarrao, Z. Fisk, A. Lacerda, C. Mielke, 
and J. D. Thompson: Phys. Rev. B {\bf 56} (1997) 71. 
\bibitem{Basu} S. Basu and P. S. Riseborogh: Physica B {\bf 378-380} (2006) 686. 
\bibitem{Dzero2} M. O. Dzero, L. P. Gor'kov, and A.~K.~Zvezdin: 
J. Phys.: Condens. Matter {\bf 12} (2000) L711.
\bibitem{Falikov} L. M. Falicov and J. C. Kimball: Phys. Rev. Lett. {\bf 22} (1969) 997. 
\bibitem{Varma} C. M. Varma: Rev. Mod. Phys. {\bf 48} (1976) 219. 
\bibitem{FK} C. E. T. Concalves da Silva and L. M. Falicov: 
Solid State Commun. {\bf 17} (1975) 1521. 
\bibitem{Hewson} A. C. Hewson and P. S. Riseborough: 
Solid State Commun. {\bf 22} (1977) 379. 
\bibitem{HF1} I. Singh,  A. K. Ahuja and S. K. Joshi: Solid State Commun. {\textbf 34} (1980) 65. 
\bibitem{Zlatic} J. K. Freericks and V. Zlatic: Phys. Rev. B {\bf 58} (1998) 322. 
\bibitem{Golstev} A. V. Goltsv and G. Bruls: Phys. Rev. B {\bf 63} (2001) 155109. 
\bibitem{Ceband} W. E. Pickett, A. J. Freeman, and D.~D.~Koelling: 
Phys. Rev. B {\bf 23} (1981) 1266. 
\bibitem{Takegahara} K. Takegahara and T. Kasuya: J. Phys. Soc. Jpn {\bf 59} (1990) 3299. 
\bibitem{Yoshikawa} K.~Yoshikawa, H.~Sato, M.~Arita, Y.~Takeda, K.~Hiraoka, K.~Kojima, K.~Tsuji, 
H.~Namatame, and M.~Taniguchi: Phys. Rev. B {\bf 72} (2005) 165106.  
\bibitem{slbMF} N. Read and D. M. Newns: J. Phys. C: Solid State Phys. {\bf 16} (1983) 3273. 
\bibitem{noteMF} Beyond the mean-field framework, the FOVT line has been shown to satisfy $dU_{\rm fc}/d\varepsilon_{\rm f}=-1$ in the $\varepsilon_{\rm f}$-$U_{\rm fc}$ plane in the large $U_{\rm fc}$ regime by the thermodynamic relation irrespective of total filling, which has been confirmed by the DMRG calculation. See ref.~\citen{WIM} for details. 
\bibitem{WM} S. Watanabe and K. Miyake: arXiv:0906.3986. 
\bibitem{Allen} J. Allen and L.~Z.~Liu.: Phys. Rev. B {\bf 46} (1992) 5047. 
\bibitem{Wohlleben}  D. Wohlleben and J. R{\" o}hler: J. Appl. Phys. {\bf 55} (1984) 15. 
\bibitem{TM} T. Miyake: private communications. 
\bibitem{WI_thermo} S. Watanabe and M. Imada: J. Phys. Soc. Jpn. {\bf 73} (2004) 3341. 
\bibitem{Dzero} M. Dzero, M.~R. Norman, I.~Paul, C.~Pepin, and J.~Schmalian: 
Phys. Rev. Lett. {\bf 97} (2006) 185701. 
\bibitem{sarraoPhysicaB1999} J. L. Sarrao: Physica B {\bf 259-261} (1999) 128. 
\bibitem{Millis2} A. J. Millis, A. J. Schofield, G.~G.~Lonzarich, and S.~A.~Grigera: 
Phys. Rev. Lett. {\bf 88} (2002) 217204. 
\bibitem{Mignot} J.-M. Mignot, J. Flouquet, P. Haen, F. Lapierre, L. Puech, and 
J. Voiron: J. Magn. Magn. Mater. {\bf 76-77} (1988) 97. 
\bibitem{sakakibara} T. Sakakibara, T. Tayama, K. Matsuhira, H. Mitamura, H.~Amitsuka, 
K.~Maezawa, and Y. $\rm \bar{O}$nuki: Phys. Rev. B {\bf 51} (1995) 12030. 
\bibitem{Aoki} Y. Aoki, T. D. Matsuda, H. Sugawara, H. Sato, H.~Ohkumi, R.~Settai, 
Y.~$\rm \bar{O}$nuki, E.~Yamamoto, Y.~Haga, A.~V.~Andreev, V.~Sechovsky, L. Havela, H. Ikeda, 
and K. Miyake: J. Magn. Magn. Mater. {\bf 177-181} (1998) 271. 
\bibitem{Evans} S. M. M. Evans: J. Magn. Magn. Mater. {\bf 108} (1992) 135.
\bibitem{Ono} Y. $\rm \bar{O}$no: J. Phys. Soc. Jpn. {\bf 67} (1998) 2197. 
\bibitem{wataKLM} S. Watanabe: J. Phys. Soc. Jpn. {\bf 69} (2000) 2947. 
\bibitem{MI} K. Miyake and H. Ikeda: J. Phys. Soc. Jpn. {\bf 75} (2006) 033704. 
\bibitem{white1} S. R. White: Phys. Rev. Lett. {\bf 69} (1992) 2863.
\bibitem{white2} S. R. White: Phys. Rev. B {\bf 48} (1993) 10345.
\bibitem{Saiga} Y. Saiga, T. Sugibayashi, and D. S. Hirashima: 
J. Phys. Soc. Jpn. {\bf 77} (2008) 114710. 
\bibitem{Knebel} G. Knebel,  R.~Boursier, E.~Hassinger, G.~Lapertot, P.~G.~Niklowitz, A.~Pourret, 
B.~Salce, J.~P.~Sanchez, I.~Sheikin, P.~Bonville, H.~Harima, and J.~Flouquet: 
J. Phys. Soc. Jpn. {\bf 75} (2006) 114709. 
\bibitem{Sarrao1996} J. L. Sarrao, C.~L.~Benton, Z.~Fisk, J.~M.~Lawrence, D.~Mandrus, and 
J.~D.~Thompson: Physica B {\bf 223\&224} (1996) 366. 
\bibitem{Cornelious} A. L. Cornelius, J. M. Lawrence, J.~L.~Sarrao, Z.~Fisk, 
M.~F.~Hundley, G.~H.~Kwei, J.~D.~Thompson, C.~H.~Booth, and F.~Bridges: 
Phys. Rev. B {\bf 56} (1997) 7993. 
\bibitem{Dallera} C. Dallera, M. Grioni, A.~Shukla, G.~Vanko, J.~L.~Sarrao, J.~P.~Rueff, 
and D.~L.~Cox: Phys. Rev. Lett. {\bf 88} (2002) 196403. 
\bibitem{Matsuda2} Y. H. Matsuda, T. Inami, K.~Ohwada, Y.~Murata, H.~Nojiri, Y.~Murakami, 
H.~Ohta, W.~Zhang, and K.~Yoshimura: J. Phys. Soc. Jpn. {\bf 76} (2007) 034702. 
\bibitem{Figueroa} E. Figueroa, J. M. Laurence, J.~L.~Sarrao, Z.~Fisk, M.~F.~Hundley, and J.~D.~Thompson: Solid State Commun. {\bf 106} 347.
\bibitem{Rossel} C. Rossel, K. N. Yang, M. B. Maple, Z. Fisk, E. Zirngiebl, and J. D. Thompson: 
Phys. Rev. B {\bf 35} (1987) 1914. 
\bibitem{Severing} A. Severing, A. P. Murani, J. D. Thompson, Z.~Fisk, and C.-K.~Loong: 
Phys. Rev. B {\bf 41} (1990) 1739. 
\bibitem{Sarrao} J. L. Sarrao, C. D. Immer, Z.~Fisk, C.~H.~Booth, E.~Figueroa, J.~M.~Lawrence, 
R.~Modler, A.~L.~Cornelius, M.~F.~Hundley, G.~H.~Kwei, J.~D.~Thompson, and F.~Bridges: 
Phys. Rev. B {\bf 59} (1999) 6855. 
\bibitem{Hiraoka} K. Hiraoka, K. Kojima, T. Hihara, and T. Shinohara: 
J. Mag. Mag. Mater. {\bf 140-144} (1995) 1243. 
\bibitem{Rice} T. M. Rice and K. Ueda: Phys. Rev. B {\bf 34} (1986) 6420. 
\bibitem{Koyama} T. Koyama, M. Matsumoto, T.~Tanaka, H.~Ishida, T.~Mito, S.~Wada, 
and J.~L.~Sarrao: Phys. Rev. B {\bf 66} (2002) 014420. 
\bibitem{Takeuchi} T. Takeuchi, T. Inoue, K.~Sugiyama, D.~Aoki, Y.~Tokiwa, Y.~Haga, K.~Kindo, 
and Y.~$\rm {\bar O}$nuki: J. Phys. Soc. Jpn. {\bf 70} (2001) 877. 
\bibitem{Kim} J. S. Kim, J. Alwood, P.~Kumar, and G.~R.~Stewart: 
Phys. Rev. B {\bf 65} (2002)  174520. 
\bibitem{Parm} E. C. Parm, T. P. Murphy, D.~Hall, S.~W.~Tozer, R.~G.~Goodrich, 
and J.~L.~Sarrao: Physica B {\bf 329-333} (2003) 587. 
\bibitem{Capan} C. Capan, A. Bianchi, F.~Ronning, A.~Lacerda, J.~D.~Thompson, M.~F.~Hundley, 
P.~G.~Pagliuso, J.~L.~Sarrao, and R.~Movshovich: 
Phys. Rev. B {\bf 70} (2004) 180502. 
\bibitem{CEF} A. D. Christianson, E.~D.~Bauer, J.~M.~Lawrence, P.~S.~Riseborough, N.~O.~Moreno, 
P.~G.~Pagliuso, J.~L.~Sarrao, J.~D.~Thompson, E.~A.~Goremychkin, F.~R.~Trouw, M.~P.~Hehlen, 
and R.~J.~McQueeney: 
Phys. Rev. B {\bf 70} (2004) 134505. 
\bibitem{Harima} Y. Haga, Y. Inada, H. Harima, K.~Oikawa, M.~Murakawa, H.~Nakawaki, 
Y.~Tokiwa, D.~Aoki, H.~Shishido, S.~Ikeda, N.~Watanabe, and Y.~$\rm {\bar O}$nuki: 
Phys. Rev. B {\bf 63} (2001) 060503. 
\bibitem{Shishido} H. Shishido, R. Settai, D.~Aoki, S.~Ikeda, H.~Nakawaki, N.~Nakamura, 
T.~Iisuka, Y.~Inada, K.~Sugiyama, T.~Takeuchi, K.~Kindo, T.~C.~Kobayashi, Y.~Haga, H.~Harima, 
Y.~Aoki, T.~Namiki, H.~Sato, and Y.~$\rm {\bar O}$nuki: 
J. Phys. Soc. Jpn. {\bf 71} (2002) 162. 
\bibitem{Hegger} H. Hegger, C. Petrovic, E. G. Moshopoulou, M.~F.~Hundley, J.~L.~Sarrao, Z.~Fisk, 
and J.~D.~Thompson: Phys. Rev. Lett. {\bf 84} (2000) 4986. 
\bibitem{Muramatsu} T.~Muramatsu, N.~Tateiwa, T.~C.~Kobayashi, K.~Shimizu, 
K.~Amaya, D.~Aoki, H.~Shishido, Y.~Haga, and Y. $\rm \bar{O}$nuki: J. Phys. Soc. Jpn. {\bf 70} (2001) 3362. 
\bibitem{Knevel2008} G. Knebel, D. Aoki, J.-P.~Brison, and J.~Flouquet: 
J. Phys. Soc. Jpn. {\bf 77} (2008) 114704. 
\bibitem{MN} K. Miyake and O. Narikiyo: J. Phys. Soc. Jpn. {\bf 71} (2002) 867. 
\bibitem{Phillips} N. E. Phillips, R. A. Fisher, F.~Bouquet, M.~F.~Hundley, 
P.~G.~Pagliuso, J.~L.~Sarrao, Z.~Fisk, and J.~D.~Thompson: 
J. Phys.: Condens. Matter {\bf 15} (2003) S2095. 
\bibitem{Park2006} T. Park, F. Ronning, H.~Q.~Yuan, M.~B.~Salamon, R.~Movshovich, 
J.~L.~Sarrao, and J.~D.~Thompson: Nature {\bf 440} (2006) 65. 
\bibitem{Shishido2}
H. Shishido, R. Settai, H. Harima, and Y. \=Onuki: 
J. Phys. Soc. Jpn. {\bf 74} (2005) 1103.
\bibitem{Sheikin}
I. Sheikin, A. Huxley, D. Braithwaite, J.-P. Brison, S. Watanabe, 
K. Miyake, and J. Flouquet: 
Phys. Rev. B {\bf 64} (2001) 220503. 
\bibitem{WatanabeUGe2}
S. Watanabe and K. Miyake: J. Phys. Soc. Jpn. {\bf 71} (2002) 2489. 
\bibitem{Park2009} T. Park, V. A. Sidorov, R. Ronning, J.-X.~Zhu, Y.~Tokiwa, H.~Lee, 
E.~D.~Bauer, R.~Movshovich, J. L. Sarrao, and J. D. Thompson: Nature {\bf 456} (2008) 366. 
\bibitem{kawasaki} S. Kawasaki, M.~Yashima, Y.~Mugino, H.~Mukuda, Y.~Kitaoka, 
H.~Shishido, and Y.~$\rm {\bar O}$nuki: 
Phys. Rev. Lett. {\bf 94} (2005)  037007. 
\bibitem{nakajima} Y. Nakajima, H.~Shishido, H.~Nakai, T.~Shibauchi, M.~Hedo, Y.~Uwatoko, 
T.~Matsumoto, R.~Settai, Y.~$\rm {\bar O}$nuki, H.~Kontani, and Y. Matsuda: 
Phys. Rev. B {\bf 77} (2008)  214504. 
\bibitem{maehira} T. Maehira, T. Hotta, K. Ueda, and A. Hasegawa: J. Phys. Soc. Jpn. {\bf 72} (2003) 854.
%
\end{thebibliography}
\end{document}